\documentclass[useAMS]{mn2e}
\twocolumn
\input{psfig}
\usepackage{times}
\usepackage{bm}
\usepackage{amsmath} 
\usepackage{url}
\usepackage{amssymb}
\usepackage{dblfnote}
\usepackage{aas_macros}
\usepackage{hyperref}
\usepackage{ifthen}
\def\Real{{\rm I\mathchoice{\kern-0.70mm}{\kern-0.70mm}{\kern-0.65mm}%
 {\kern-0.50mm}R}}

\font \bolditalics = cmmib10
\def\bx#1{\leavevmode\thinspace\hbox{vrule\vtop{\vbox{\hrule\kern1pt
 \hbox{\vphantom{\tt/}\thinspace{\bf#1}\thinspace}}
 \kern1pt\hrule}\vrule}\thinspace}

\def \vc #1{{\textfont1=\bolditalics \hbox{$\bf#1$}}}

\def\sg{\bmath{s}}
\def\xg{\bmath{x}}

\def\kg{\bmath{k}}

\def\tg{ \btheta}

\def\gg{{\vc \gamma}}

\def\be{\begin{equation}}
\def\ee{\end{equation}}

\def\ba{\begin{eqnarray}}
\def\ea{\end{eqnarray}}

\def \vc #1{{\textfont1=\bolditalics \hbox{$\bf#1$}}}{\catcode`\@=11

\def\ave#1{\left\langle #1 \right\rangle}

\def\ltsima{$\; \buildrel < \over \sim \;$}
\def\lsim{\lower.5ex\hbox{\ltsima}}
\def\gtsima{$\; \buildrel > \over \sim \;$}
\def\gsim{\lower.5ex\hbox{\gtsima}}

\title[Baryonic effect on two- and three-point shear statistics]{Effect of  baryonic feedback on two- and three-point shear statistics: prospects for detection and improved modelling}

\author[Semboloni et al.]{Elisabetta Semboloni$^{1}$\thanks{sembolon@strw.leidenuniv.nl}, Henk Hoekstra$^1$ \& Joop Schaye$^1$  \\
$^1$Leiden Observatory, Leiden University, P.O. Box 9513, NL 2300 RA, Leiden, The Netherlands \\
}
\begin{document}
\maketitle
\begin{abstract}

Accurate knowledge of the effect of feedback from galaxy formation on the matter distribution is a key requirement for future weak lensing experiments. Recent studies using hydrodynamic simulations have shown that different baryonic feedback scenarios lead to significantly different two-point shear statistics.  In this paper we extend earlier work  to three-point shear statistics. 
We show that, relative to the predictions of dark matter only models, the amplitude of the signal can be reduced by as much as $30-40\%$ on scales of a few arcminutes.  As is the case for two-point shear tomography, the interpretation of three-point shear statistics with dark matter only models is therefore plagued by a strong bias. 
However, we find that baryonic feedback  may affect two- and three-point shear statistics differently and demonstrate that this can be used to assess the fidelity of various feedback models. In particular, upcoming surveys such as {\it Euclid}   might be able to discriminate between different feedback models by measuring both second- and third-order shear statistics. Because it will likely remain impossible to predict baryonic feedback with high accuracy from first principles, we argue in favour of phenomenological models that can capture the relevant effects of baryonic feedback processes in addition to changes in cosmology. 
We construct such a model by modifying the standard (dark matter only) halo model to characterise the generic effects of energetic feedback using a small number of parameters.  We use this model to perform a likelihood analysis in a simplified case in which  two- and three-point shear statistics are measured over the range $0.5<\theta<20$ arcmin and in which the amplitude of fluctuations, $\sigma_8$, the matter density parameter, $\Omega_m$, and the dark energy parameter, $w_0$, are the only unknown free parameters.
  We demonstrate  that for weak lensing surveys such as {\it Euclid},  marginalising over the feedback parameters describing the effects of baryonic processes, such as outflows driven by feedback from star formation and AGN, may be able to mitigate the bias affecting  $\Omega_m$, $\sigma_8$ and $w_0$. 
\end{abstract}


\section{Introduction}
Weak lensing by large-scale structure, i.e.\ cosmic shear, is sensitive to the matter distribution without having to rely on assumptions about its nature or dynamic state. Thus, it represents a powerful tool to study  the statistical properties of matter density fluctuations. In particular,  the measurement of the  two-point shear statistics as a function of the redshift of the sources, allows  one to study the evolution of the power spectrum of the  matter fluctuations  (e.g.\ Hu 2002). In this respect, weak lensing is complementary to primary cosmic microwave background (CMB) experiments, which by themselves do not  constrain the late-time evolution of density fluctuations. For a recent review of cosmological applications, see Hoekstra \& Jain (2008).

 The success of pioneering measurements led to dedicated surveys such as the Canada France Hawaii Legacy Survey (CFHTLS), the ongoing Kilo-Degree Survey (KiDS) and the Dark Energy Survey (DES). Future surveys such as the  Hyper Suprime-Cam (HSC) survey, the Large Synoptic Survey Telescope (LSST)  and {\it Euclid},{\footnote {http://www.euclid-ec.org/}}  \cite{redbook,Amendolaetal12}, will offer the possibility to test the validity of the standard cosmological model and, in particular, to discriminate  between dark energy and modified gravity models.  They will have the statistical power to constrain the equation of state of the dark energy at the percent level. However, a change of a few percent in the equation of state corresponds to a change of a percent in the amplitude of the matter power spectrum. This means that in order to exploit the statistical power of those datasets,  one needs to predict the power spectrum with percent level accuracy up to scales of $k \sim 10~h/{\rm Mpc}$ for  which  the evolution of the fluctuations is in the non-linear regime (e.g. Hearin et al. 2012).

Currently, the interpretation of two-point shear statistics relies on predictions which are based on N-body simulations of collisionless particles. To date, this has been  sufficient, as the matter present in the Universe is mainly  collisionless and statistical uncertainties large.  However, baryons represent about $17\%$  of the total matter content in the Universe \cite{WMAP7} and the  power spectrum of density fluctuations therefore  also  depends  on their distribution. 

The distribution of baryons is governed by  physical mechanisms such as radiative cooling, star formation and feedback from supernovae  and active galactic nuclei (AGN).  Those mechanisms  have been implemented only recently in hydrodynamic simulations  and the accuracy of the results is still under discussion.  Springel (2010a, 2010b, 2011)  thoroughly  discuss the performance  of the smoothed particle hydrodynamics (SPH), adaptive mesh refinement (AMR) and moving mesh methods.   Along these lines, Vogelsberger et al.\ (2011) demonstrated that different numerical techniques can lead to different properties on galactic scales. However, Schaye et al.\ (2010) and Scannapieco et al.\ (2012) suggest that the differences between  numerical techniques are smaller than the differences due  to variations  in the physical  mechanisms when considering the properties of groups.  In particular, McCarthy et al.\ (2010), Puchwein et al.\ (2010), Fabjan et al.\ (2010), and Dubois et al.\ (2011) have shown that the presence of AGNs significantly affects the properties of groups, leading to much improved agreement with observations. 

Clearly, an accurate description of the matter distribution in the Universe cannot be achieved without a correct description of the baryonic component.  Indeed, van Daalen et al.\ (2011) showed that baryonic feedback processes modify the power spectrum of matter fluctuations (see also Jing et al. 2006; Levine \&
Gnedin (2006); Rudd et al. 2008; Guillet et al.
2010; Casarini et al. 2011) and that, depending on the scenario, this modification can affect  surprisingly large scales. In particular, van Daalen et al.\ (2011) found that in their model with AGN feedback, their only model for which predictions for groups are in good agreement with both X-ray and optical observations (McCarthy et al.\ 2010), the matter power spectrum is suppressed by 1 per cent and 10 per cent at $k\approx 0.3$ and $1~h/{\rm Mpc}$, respectively. Semboloni et al.\ (2011a) used those same power spectra to derive the cosmic shear signal for  different scenarios and  showed that ignoring baryonic feedback when interpreting two-point shear statistics can strongly bias the measurements of cosmological parameters.  While these studies focused on two-point statistics, higher order statistics must also be affected.
For instance, Yang et al. (2013) recently investigated how  the presence of baryons affects peak statistics. For their study, they tried to reproduce the steepening  of the halo density profiles in the central regions caused by star formation,  by  manually increasing the halo concentrations of dark matter only simulations. They find a mild increase of high peaks, but the effect  could be different when using  hydrodynamic simulations.  
  
In this paper we will investigate the effect of baryonic feedback on three-point shear statistics, which have been advocated in the past as an additional cosmological probe. For instance, previous studies  have shown that the combination of two- and three-point shear statistics provides much tighter cosmological constraints than the two-point shear statistics  alone \cite{Beetal97,TaJa03a,KiSc05,Beetal09,Vaetal10,Kayoetal13}. 
 The benefit of combining three- and  two-point shear statistics  is even larger in the self-calibration regime \cite{Huetal06}, where one exploits the different sensitivity  of the two measurements  to the source redshift distribution in order  to alleviate the impact  of the lack of accurate knowledge of the redshifts of the source galaxies.

A few attempts have been made to measure three-point shear statistics \cite{Peetal03,JaBeJa04}. Those early  measurements  used small  datasets  and  shear measurement  algorithms which were much less accurate than the ones available today. While in both studies a signal was detected, the results were affected by residual point spread function (PSF) systematics. More recently,  Semboloni et al.\ (2011b)  used  high-quality space-based  data, the COSMOS-HST dataset, to perform a measurement of  three-point shear statistics  which does not show  evidence of residual systematics.

 For large datasets  from ongoing and future surveys the statistical power of  three-point shear statistics is comparable to that of two-point shear statistics  \cite{Vaetal10}. Thanks to its different dependency on the cosmological parameters, three-point shear statistics will thus  be a complementary test for  cosmological models.
However, the enhanced sensitivity to the  non-linear features of the field of  density fluctuations  makes  the interpretation  of three-point shear statistics  challenging. To date, this interpretation has relied on models  based either on perturbation theory \cite{ScFr99,ScCo01} or on the halo model \cite{TaJa03a,TaJa03b}.  These  models have only been tested against dark matter  simulations and do not agree to better than $10-20\%$. Moreover,  the effect of baryonic physics on three-point shear statistics,  or equivalently, on the bispectrum of matter fluctuations  is still  unknown. 

In this paper we use hydrodynamic simulations from the OverWhelmingly Large Simulations (OWLS; Schaye et al. 2010)  to assess  the impact of baryonic feedback on higher order statistics. The paper is organised as follows:  in Section \ref{sec:methode} we describe the method used to compute three-point shear statistics;  in Section \ref{sec:sims} we  briefly describe the simulations; in Section \ref{sec:results} we present the results from the OWLS models. In Section \ref{sec:newmodel} we introduce a new parametrisation for the effect of baryonic feedback on the matter power spectrum which updates the one by Semboloni et al.\ (2011a). In  Section \ref{sec:likelihood} we perform a likelihood analysis showing how the simultaneous measurement of the two- and three-point shear statistics can detect the presence of feedback.  In this way, three-point shear statistics can be used to self-calibrate for the (unknown) baryonic feedback.
Furthermore, we show how  a general parametrisation of the effect of feedback, the strength of which is only marginally constrained, can be used to reduce the effect of the bias on the cosmological parameter constraints. Finally, we discuss our results and conclude in Section \ref{sec:conclusions}. 

\section{Calculation of three-point shear statistics}\label{sec:methode}
In this section we recall briefly the relation between cosmic shear  and  density fluctuation statistics and introduce the notation used in this paper. We define the matter density contrast $\delta$  such that  the density field at a position $\xg$ is $\rho(\xg)\equiv \bar \rho (1+\delta(\xg))$, with $\bar \rho$ the average density. We define the power spectrum $P(k)$ and the bispectrum $ B(k_1,k_2,k_3)$ as:
{\setlength\arraycolsep{0.05em}
\begin{eqnarray}
\ave{\tilde\delta(\kg_1)\tilde\delta(\kg_2)}&\equiv&(2\pi)^3\delta_{\rm D}(\kg_1+\kg_2) P(k_1)\label{eq:spectrum},\\ 
\ave{\tilde\delta(\kg_1)\tilde\delta(\kg_2)\tilde\delta(\kg_3)}&\equiv&(2\pi)^3\delta_{\rm D}(\kg_1+\kg_2+\kg_3) B(k_1,k_2,k_3)\,\label{eq:bispectrum},
\end{eqnarray}
}
\noindent  where $\tilde\delta$ is the Fourier transform of the density contrast $\delta$ and $\kg$ is the Fourier conjugate of $\xg$. The fact that $\ave{\tilde\delta(\kg_1)\tilde\delta(\kg_2)}$  differs from zero only if $\kg_1=-\kg_2$ and depends only on the modules of the wave vector is a consequence of assuming $\delta$ to be a statistically homogeneous and isotropic field. This also implies that $\ave{\tilde\delta(\kg_1)\tilde\delta(\kg_2)\tilde\delta(\kg_3)}$ is defined only for closed triangles and does not depend on their orientation.  

The effect of foreground overdensities on  background galaxies is a stretching (i.e.\ a circular source would appear elliptical) and a (de)magnification. Those effects are completely  described  by the complex  shear $\gg=\gamma_1+ i\gamma_2$  and the convergence $\kappa$. Both are related to the second order derivatives of the projected gravitational potential (see for example Schneider et al. 1998). In particular:
\be\label{eq:kappa}
\kappa(\sg)=\frac{3}{2}\Big(\frac{H_0}{c}\Big)^2 \Omega_m\int_0^{w_h} dw g(w)f_k(w)\frac{\delta(\sg/f_k(w),w)}{a(w)}\,,
\ee
\noindent where $\sg$ is a two-dimensional vector such that $\sg=\kg_\perp  f_k(w)$ with $\kg_\perp$ the projection of $\kg$ perpendicular to the line of sight. $H_0$ is the Hubble constant,  $\Omega_m$ is the matter density parameter, $c$ is the speed of light,  $w$ is the comoving radial coordinate, $f_k(w)$ is the comoving angular diameter distance at radial distance $w$, $a$ is the cosmological scale factor,
\be\label{eq:redshift}
g(w)=\int_w^{w_H} p_w(w^\prime) d w^\prime \frac{f_k(w^\prime -w)}{f_k(w^\prime)}\,,
\ee
\noindent and $p_w(w)dw$ describes the distribution of sources as a function of the radial coordinate, or equivalently as a function of the redshift, in which case it is  denoted as $p_z(z)dz$.

Analogous to Equations (\ref{eq:spectrum}) and (\ref{eq:bispectrum}), we define the convergence power spectrum and bispectrum:
{\setlength\arraycolsep{0.1em}
\ba
\ave{\tilde\kappa(\sg_1)\tilde\kappa(\sg_2)}&\equiv&(2\pi)^2\delta_{\rm D}(\sg_1+\sg_2) P_\kappa(s)\, ,\\
\ave{\tilde\kappa(\sg_1)\tilde\kappa(\sg_2)\tilde\kappa(\sg_3)}&\equiv&(2\pi)^2\delta_{\rm D}(\sg_1+\sg_2+\sg_3) B_\kappa(s_1,s_2,s_3)\,.
\ea
}
\noindent One cannot measure $\kappa$  directly, but one can instead estimate the shear $\gamma$ by measuring the ellipticities of galaxies. In fact,  assuming  that galaxies are randomly oriented,  any  observed correlation  between the orientations of galaxies is the result of the deflection of light  by  foreground structures and thus a direct measurement of the shear correlations. The power spectra of $\kappa$ and $\gamma$ are the same and the relation between two-point shear correlations and $P_\kappa(s)$ is straightforward.

 The relation between the three-point shear correlation functions  and the bispectrum is more complicated \cite{ScKiLo05}. For this reason,  the measured three-point shear correlation functions are generally combined  to obtain a new third-order statistic:
{\setlength\arraycolsep{0.05em}
\ba\label{eq:map3}
\ave{M_{\rm ap}^3(\theta)}=\int d^2\theta_1 U_\theta( \theta_1)&& \int d^2\theta_2 U_\theta( \theta_2) \int d^2\theta_3 U_\theta( \theta_3)\\
\nonumber \int\frac{d^2 s_1 e^{i \tg_1 \sg_1}}{(2 \pi)^2}\int\frac{d^2 s_2  e^{i \tg_2 \sg_2}}{(2 \pi)^2} &&\int\frac{d^2 s_3 e^{i \tg_3 \sg_3}}{(2 \pi)^2}\ave{\tilde\kappa(\sg_1)\tilde\kappa(\sg_2)\tilde\kappa(\sg_3)}
\ea
}

\noindent which is directly related to the projected bispectrum. In fact, by choosing
\noindent 
\be\label{eq:filter}
U_\theta(\vartheta)=\frac{1}{2\pi \theta^2}\Big(1-\frac{\vartheta^2}{2 \theta^2} \Big) e^{-\frac{1}{2}(\frac{\vartheta}{\theta})^2}\,,
\ee
\noindent the relation between three-point correlation functions and $\ave{M_{\rm ap}^3(\theta)}$ is relatively simple \cite{Peetal03,JaBeJa04,ScKiLo05}.  For completeness, we remind the reader that:
\ba
\ave{M_{\rm ap}^2(\theta)}= \int d^2\theta_1 U_\theta( \theta_1) \int d^2\theta_2 U_\theta( \theta_2)\\
\int\frac{d^2 s_1 e^{i \tg_1 \sg_1}}{(2 \pi)^2} \int\frac{d^2 s_2  e^{i \tg_2 \sg_2}}{(2 \pi)^2}\ave{\tilde\kappa(\sg_1)\tilde\kappa(\sg_2)}\nonumber\,.
\ea

 Inserting Equations (\ref{eq:kappa}) and (\ref{eq:redshift})  into Equation (\ref{eq:map3}) and using the definition in Equation (\ref{eq:bispectrum}), one derives an explicit relation between the 3D bispectrum of density fluctuations and  $\ave{M_{\rm ap}^3(\theta)}$. Thus, to measure  $\ave{M_{\rm ap}^3(\theta)}$ for various sets of simulations, we in principle need to estimate the bispectrum. Measuring the 3D bispectrum directly in each  simulation  as a function of redshift is very time consuming  and we leave this for  future work. Here, we make use of an approximation  derived using perturbation theory \cite{Fry84} and adapted to cold dark matter (CDM) cosmologies by Scoccimarro \& Frieman (1999)  and Scoccimarro \& Couchman (2001):
\be\label{eq:approx_from}
 B(k_1,k_2,k_3)=2 F(\kg_1,\kg_2) P(k_1)P(k_2)+  cycl.
\ee 
\noindent where $cycl.$ indicates a cyclic permutation of the indices,
{\setlength\arraycolsep{0.1em}
\ba
F(\kg_1,\kg_2)&=&\frac{5}{7}a(n,k_1)a(n,k_2)\\
&+&\frac{1}{2}\frac{\kg_1\kg_2}{k_1k_2}\Big(\frac{k_1}{k_2}+ \frac{k_2}{k_1}\Big)b(n,k_1)b(n,k_2)\nonumber\\
&+&\frac{2}{7}\Big(\frac{\kg_1\kg_2}{k_1k_2}\Big)^2c(n,k_1)c(n,k_2)\,.\nonumber
\ea
}
\noindent where $n$ is the spectral index of the primordial power spectrum. The coefficients
\ba
a(n,k)&=&\frac{1+\sigma_8(z)^{-0.2}[0.7 Q_3(n)]^{1/2}(q/4)^{n+3.5}}{1+(q/4)^{n+3.5}}\,,\\
b(n,k)&=&\frac{1+0.4(n+0.3)q^{n+3}}{1+q^{n+3.5}}\,,\\
c(n,k)&=&\frac{1+4.5/[1.5+(n+3)^4](2q)^{n+3}}{1+(2q)^{n+3.5}}\,,
\ea
\noindent have been fitted to N-body simulations with
\be
Q_3(n)=\frac{4-2^n}{1+2^{n+1}}\label{eq:approx_to}\,.
\ee
$\sigma_8(z)$ is the standard deviation of  density fluctuations linearly evolved to redshift $z$, integrated in a sphere of radius $8~ {\rm Mpc}/h$.  The variable $q= k/k_{nl}$ with $4 \pi k_{\rm nl}^3 P_{\rm linear}(k_{\rm nl})=1$ defines a remapping. Scoccimarro \& Couchman (2001) showed that this approximation is accurate to within $15\%$  up to $k$ of a few $h/{\rm Mpc}$.

Semboloni et al.\ (2011b) compared   $\ave{M_{\rm ap}^3(\theta)}$  results from dark matter only simulations to the above  approximation  and the non-linear  halofit \cite{Smetal03}  power spectrum. They found that  $\ave{M_{\rm ap}^3(\theta)}$  is systematically underestimated on small scales, although the underestimation may be caused by the fact that  the halofit  approach underestimates the power spectrum at small scales (e.g. Hilbert et al. 2009, Semboloni et al. 2011b) and not necessarily  by the perturbative expansion.  The expression for $F(k_1,k_2)$ introduced above has been derived for CDM simulations;  the coefficients $a(n,k)$, $b(n,k)$ and $c(n,k)$ describe the transition from the quasi-linear regime, described by perturbation theory (PT), to the strongly non-linear regime. In the quasi-linear regime  ($a=b=c=1$),  we expect this formula to be valid even in the presence of a feedback. In the non-linear regime,  we  expect  $F(k_1,k_2)$ to vary as a function of the feedback scenario. However, we will assume here that  $F(k_1,k_2)$ is independent of  baryonic feedback and in doing so, we might underestimate the effect of  feedback.  In future work we will investigate the accuracy of the approximations used in this paper by directly measuring  the bispectrum in the simulations.

\section{Simulations}\label{sec:sims}
\begin{figure*}
\begin{tabular}{|@{}l@{}|@{}l@{}|@{}l@{}|} 
\psfig{figure=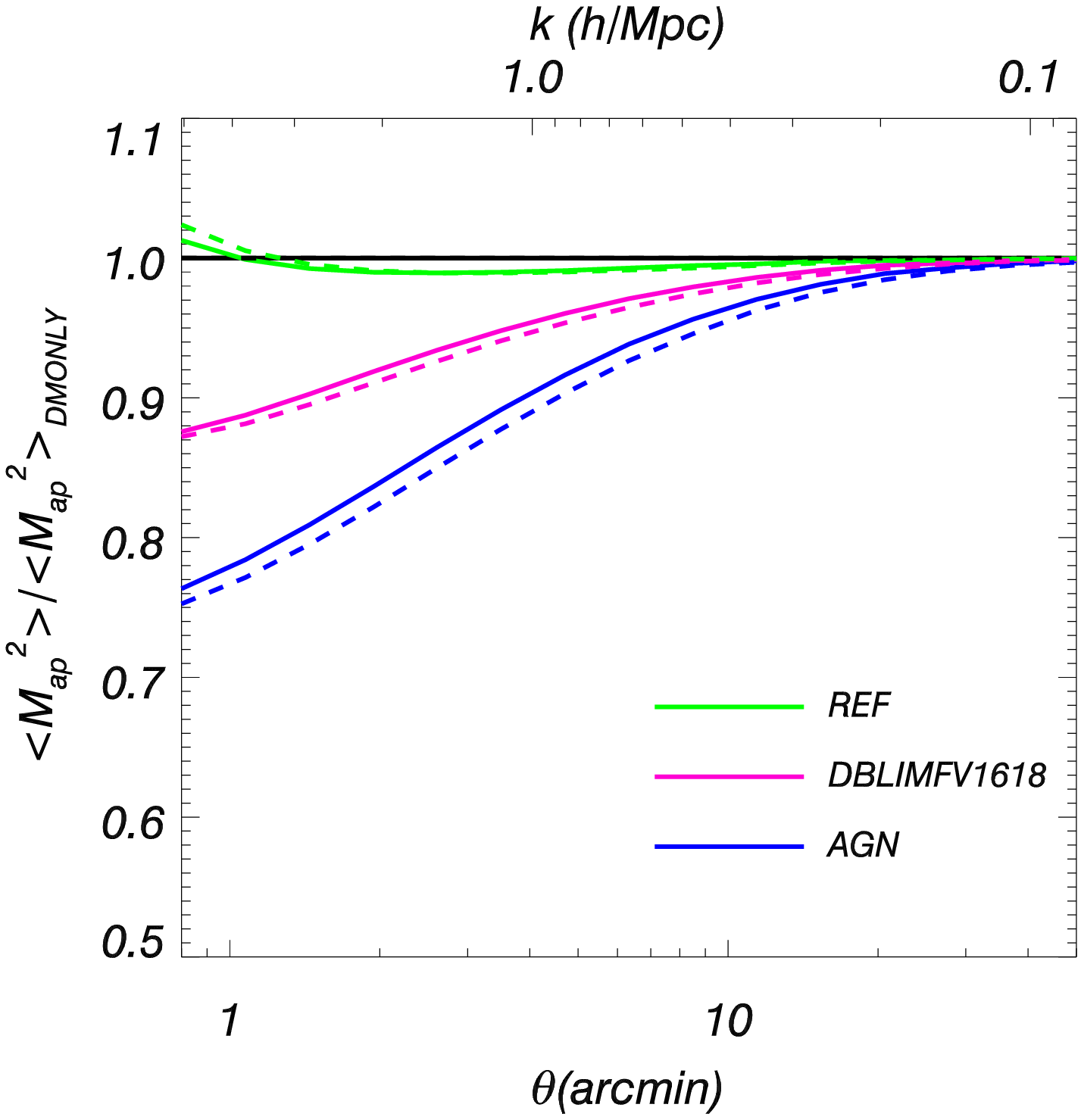,width=.33\textwidth}\psfig{figure=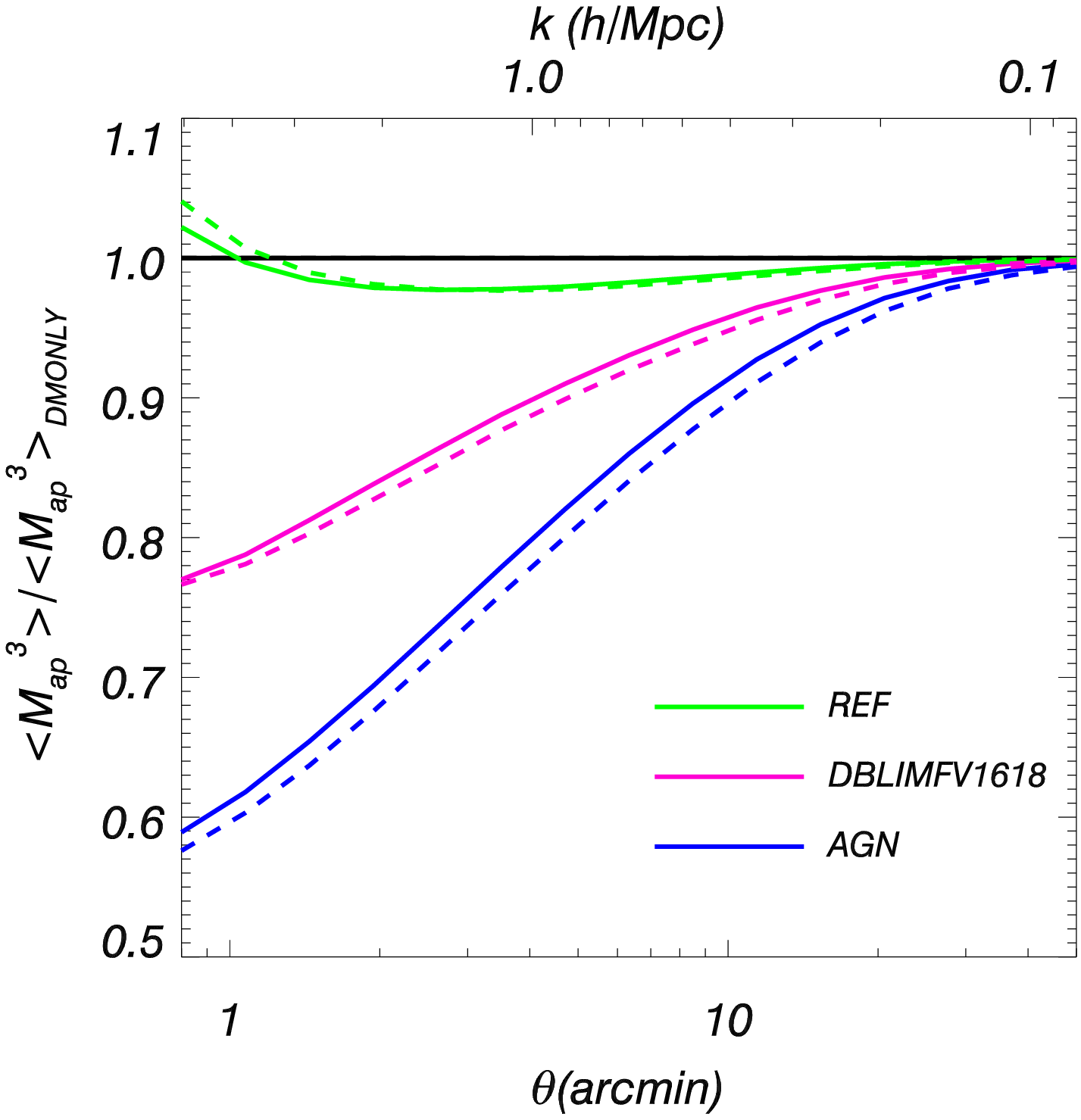,width=.33\textwidth}\psfig{figure=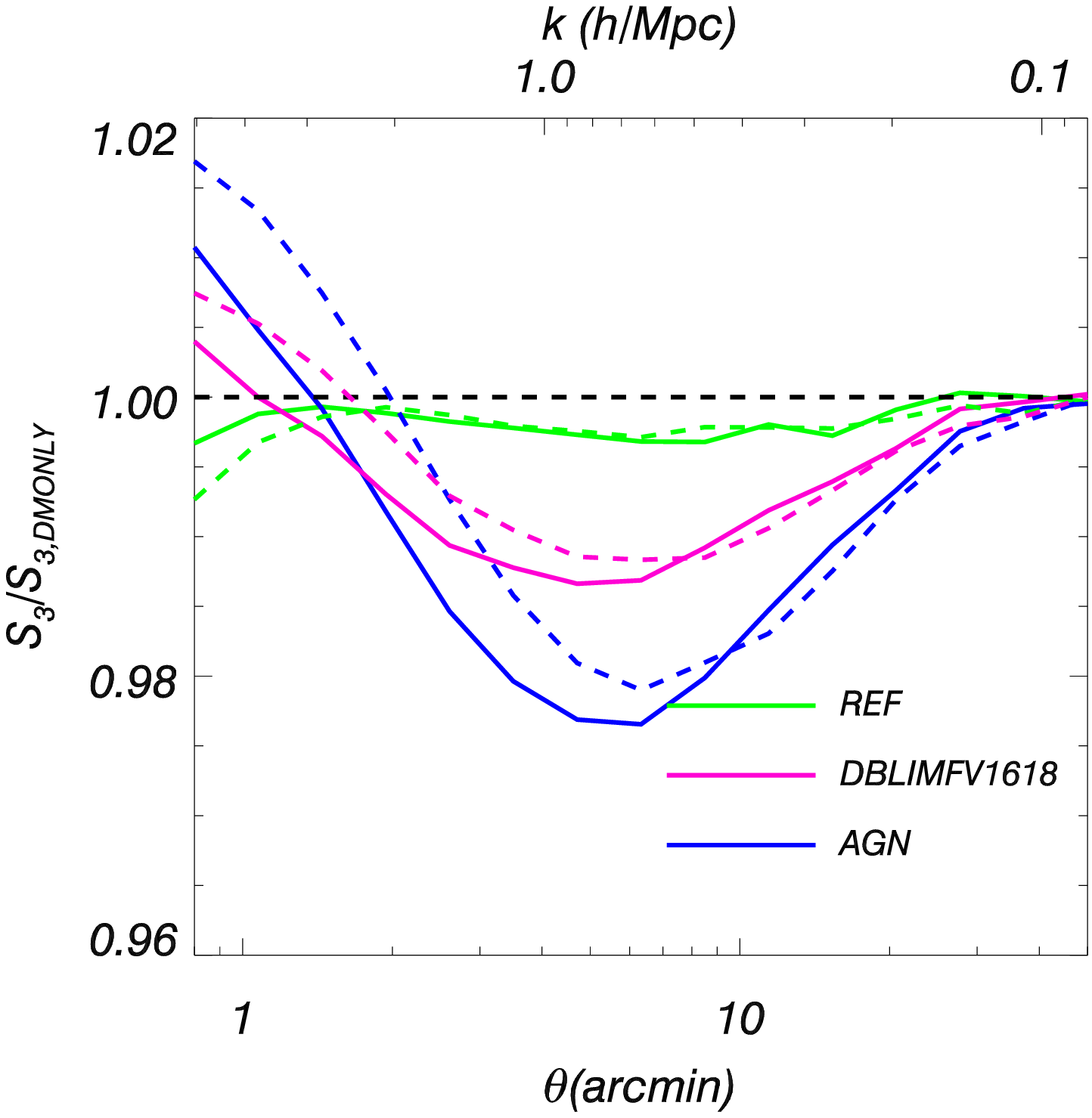,width=.33\textwidth}
\end{tabular}
\caption{\label{fig:ratio_res} The left panel shows  the ratio of $\ave{M_{\rm ap}^2(\theta)}$ of the REF (green solid line), DBLIMV1618 (pink solid line) and AGN (blue solid line) results and the DMONLY values  as a function of the angular scale $\theta$ for a survey with a similar depth as {\it Euclid} . The dashed lines show the same results for a KiDS-like survey.  The middle and right panels show the same as the left panel  but for  $\ave{M_{\rm ap}^3(\theta)}$ and $S_3(\theta)$, respectively.  We indicate on  the top the physical scale $k$ that   contributes most to the value of  $\ave{M_{\rm ap}^2(\theta)}$  and $\ave{M_{\rm ap}^3(\theta)}$ assuming a source redshift  of  $z=1$. }
\end{figure*}  
We focus here  on the four simulations  from   the OverWhelmingly Large Simulation (OWLS) project \cite{Schayeetal10} that were also analysed in Semboloni et al.\ (2011a): DMONLY, REF, DBLIMFV1618 and AGN. All the simulations have been realised using the same initial conditions  from a ${\rm \Lambda CDM}$ WMAP3 \cite{WMAP3} fiducial cosmology: $\{\Omega_m,\Omega_b,\Omega_\Lambda,\sigma_8,n,h\}=\{0.238,0.0418,0.762,0.74,0.951,0.73\}$.
The length of the box side is  $L=100 ~{\rm Mpc}~ h^{-1}$.  The DMONLY simulation contains $512^3$ collisionless particles with  mass $4.1 \times 10^8 ~M_{\odot} ~h^{-1}$. The hydrodynamic simulations contain the same number of collisionless particles plus an equal number of baryonic particles with initial mass $8.7 \times 10^7 ~M_{\odot} ~h^{-1}$.   We  describe here the other main features  of these simulations.
\begin{itemize}
\item DMONLY: a dark matter only simulation, of the kind
   commonly used to compute the non-linear power spectrum  for weak lensing studies. It is therefore the reference to which we
  compare the other simulations. 
\item REF:  this simulation represents a standard
  scenario assumed in cosmological hydrodynamic simulations.
It includes most of the mechanisms which
  are thought to be important for the star formation history (see Schaye et al. 2010 for a detailed discussion) except for AGN feedback. The implementations of radiative cooling, star formation, supernova driven winds, stellar evolution and mass loss have been described in Wiersma Schaye \& Smith (2009), Schaye \& Dalla Vecchia (2008), Dalla Vecchia \& Schaye (2008), and Wiersma et al.\ (2009) respectively. 
\item DBLIMFV1618: this simulation has been produced using the same mechanisms as REF.  The only difference between the two simulations is that  in this simulation the stellar initial mass function (IMF) 
  was modified to produce more massive stars when the pressure of the gas
  is high, i.e.\ in starburst galaxies and close to galactic
  centres.  This is accomplished by switching from the Chabrier (2003) IMF assumed in the REF model to a Baugh et al.\ (2005) IMF in those regions. The IMF change increases the number of supernovae  and thus enhances the effect of stellar winds suppressing the star formation rate (SFR) at small redshifts.   However, this mechanism alone is not able to
  reproduce the observed star formation history  (see Schaye et al. 2010).
\item AGN:  a hydrodynamic simulation which differs from REF only by the inclusion of AGN. 
 The AGN feedback
  has been modelled using a modified version of the prescription of Springel et al.\ (2005)  as described  in Booth \& Schaye (2009). In this approach
the  AGN transfers energy to the neighbouring gas, heating it up and
  driving supersonic outflows which are able to displace a large
  quantity of baryons far from the AGN itself.
\end{itemize}
In addition, the models WDENS,   NOSN\_NOZCOOL and NOZCOOL, which are described below,  are used   in Section \ref{sec:newmodel} to test the generality of our feedback model, but   their impact on the cosmic shear signal is not evaluated.
\begin{itemize}
\item WDENS:  a hydrodynamic simulation which differs from REF in that the  initial mass loading and  velocity of the supernova driven wind vary as a function the surrounding gas density so that they become more efficient in high pressure regions. 
\item NOSN\_NOZCOOL: The supernova feedback is removed and the radiative cooling rate is computed  assuming primordial abundances. 
\item NOZCOOL: Cooling rate assumes primordial abundances.
\end{itemize}
We know from  van Daalen et al.\ (2011) that the modifications to the power spectrum of the WDENS simulation are quite close to the ones of DBLIMFV1618 thus  we expect deviations in the cosmic shear signal  which are also similar.  The  NOSN\_NOZCOOL and NOZCOOL scenarios are unrealistic and/or  cannot reproduce  realistic star formation history. The power spectra in these last two simulations are not modified at large scales (see van Daalen et al.\ 2011), thus we expect changes in the cosmic shear signal similar to the ones of the REF scenario. 

Among the hydrodynamic   simulations considered here, the AGN model is arguably the most realistic, as
  it is able to reproduce the gas density, temperature, entropy, and
  metallicity profiles inferred from X-ray observations, as well as
  the stellar masses, star formation rates, and stellar age
  distributions inferred from optical observations of low-redshift
  groups of galaxies \cite{MCetal10}.

\section{Results}\label{sec:results}
%\subsection{Feedback effect}\label{subsec:results}
In this section we evaluate the effect of different baryonic feedback models on the predicted value of   $\ave{M_{\rm ap}^3(\theta)}$. We use the matter power spectra  $P(k,z)$ measured by van Daalen et al.\ (2011) together with the approximations (\ref{eq:approx_from})-(\ref{eq:approx_to}) to estimate the bispectrum. Finally, we integrate the bispectrum over the source distribution using Equations (\ref{eq:kappa}), (\ref{eq:redshift}) and (\ref{eq:map3}).
We choose the following source distribution: 
\be\label{eq:pofz}
p_z(z)=\frac{\beta_z}{z_0\Gamma(\frac{1+\alpha_z}{\beta_z})} \Big(\frac{z}{z_0}\Big)^{\alpha_z} e^{-(z/z_0)^{\beta_z}}\,,
\ee
where $\alpha_z,\beta_z$ and $z_0$ have been fixed to simulate a  KiDS-like survey and a {\it Euclid}-like survey.   Following  Vafaei et al.\ (2010),   we use the following values: $\alpha_z=0.81$, $\beta_z=3.15$, $z_0=1.19$ for the KiDS-like survey and  $\alpha_z=0.96$, $\beta_z=1.70$, $z_0=1.07$ for the {\it Euclid}-like survey, corresponding to  median redshifts of $0.8$ and $0.91$ respectively.

We show in the left panel of  Figure  \ref{fig:ratio_res} the  ratios of $\ave{M_{\rm ap}^2(\theta)}$  for REF, DBLIMFV1618 and AGN to   $\ave{M_{\rm ap}^2(\theta)}$ for DMONLY as a function of the angular scale $\theta$. Note that the aperture-mass filter defined by Equation (\ref{eq:filter})  is fairly broad and for a given angular scale $\theta$ peaks  at $s=\sqrt{2}/\theta$ (where $\theta$ is given in radians). As a consequence, aperture mass statistics  depend strongly on the density field at small scales.  As an  example, for a typical weak lensing survey, the value of  $\ave{M_{\rm ap}^2(\theta)}$ at $\theta \sim 1\, {\rm arcmin}$  depends on the power spectrum $P(k)$  at $k$ of a few  $h/{\rm Mpc}$ (see Figure 2 of Semboloni et al.\ 2011a).  Similar to what was found by Semboloni et al.\ (2011a), the suppression is in this case  about $20\%$ at arcminute scales  for the AGN scenario, whereas it is much smaller for the REF scenario.   The DBLIMFV1618 is an intermediate case  between the REF and AGN cases.

We show in the middle panel of  Figure \ref{fig:ratio_res} the ratios for  $\ave{M_{\rm ap}^3(\theta)}$.  The ratio is close to unity for the REF scenario, whereas  it has a large-scale  dependence in the case of AGN feedback, and the amplitude of  $\ave{M_{\rm ap}^3(\theta)}$ is suppressed as much as  $40\%$ on arcminute scales. Moreover, the impact is significant on large scales, being  about $10\%$ at $\theta \sim 10 ~ {\rm arcmin}$.  

Finally,  the right panel of Figure \ref{fig:ratio_res} shows the effect of baryonic feedback  on  the skewness:
\be
S_3(\theta)\equiv\frac{\ave{M_{\rm ap}^3(\theta)}}{\ave{M_{\rm ap}^2(\theta)}^2}\,.
\ee
\noindent   The skewness has been  considered a very appealing measurement as it scales  $\propto \Omega_m ^{-1}$ and depends only weakly on $\sigma_8$. In principle, it is an effective  measure to break the parameter degeneracy typical of two-point shear statistics \cite{Beetal97}. However, it has  recently been shown  \cite{Vaetal10} that  it is it much less effective to constrain cosmology even when combined with  two-point shear statistics. Indeed, the skewness has a fairly flat profile  at small scales  where the two- and three-point shear statistics show features. As a result, these statistics are more sensitive to cosmology. The right panel of Figure \ref{fig:ratio_res} shows  that the skewness is also relatively  insensitive to the baryonic feedback scenario (note the smaller y-axis  range). One of  the goals of this paper is to investigate whether  one can detect the effect of  feedback by measuring weak lensing statistics. For this reason, we focus here on the  measurement of $\ave {M_{\rm ap}^3(\theta)}$, which is more sensitive to  variations in the  feedback  model than  the skewness.

\section{Towards a general description of feedback}\label{sec:newmodel}
\subsection{Halo model and  baryonic feedback}\label{subsec:halo_model}
\begin{figure*}
\begin{tabular}{|@{}l@{}|@{}l@{}|@{}l@{}|} 
\psfig{figure=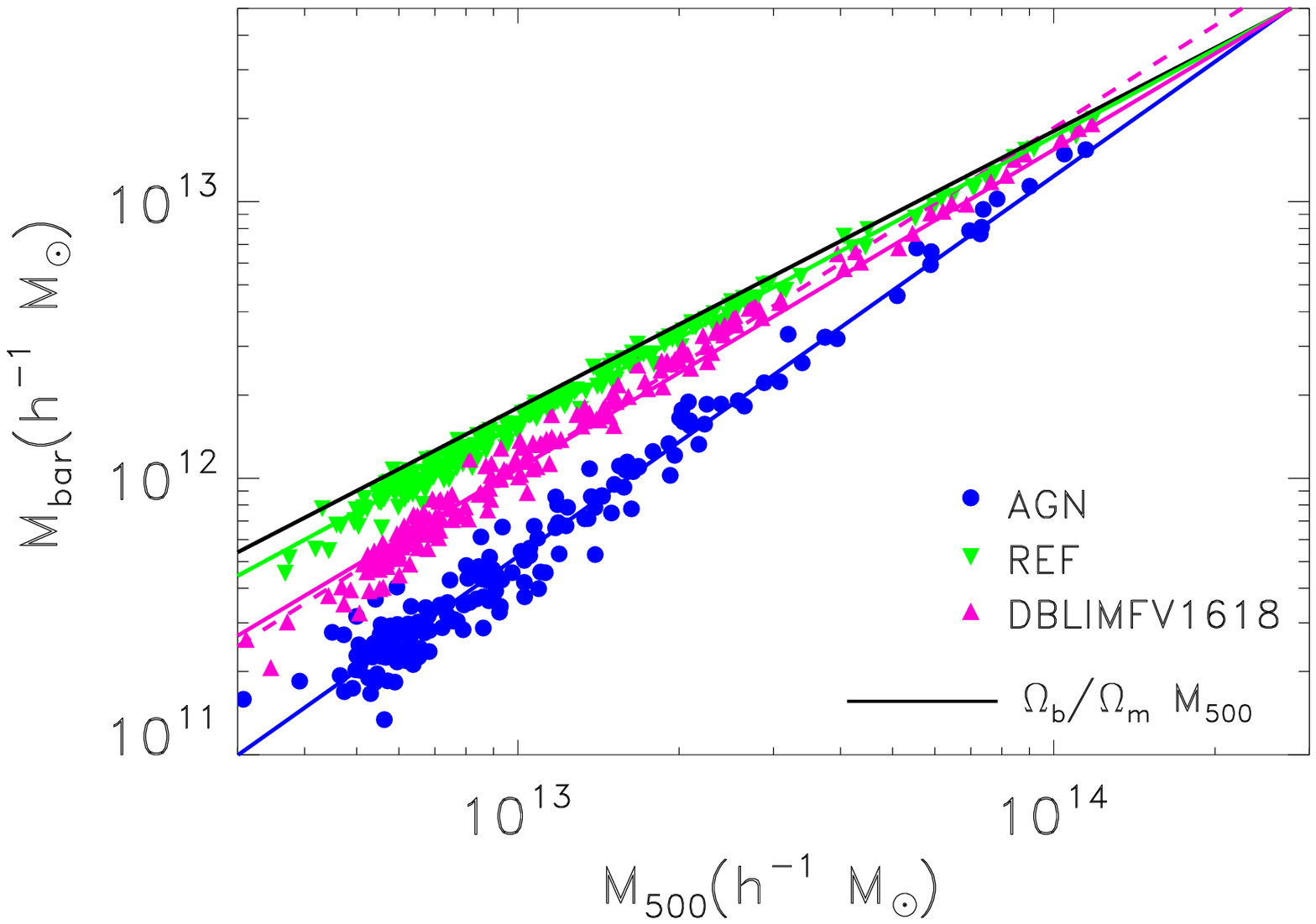,width=.50\textwidth}&\psfig{figure=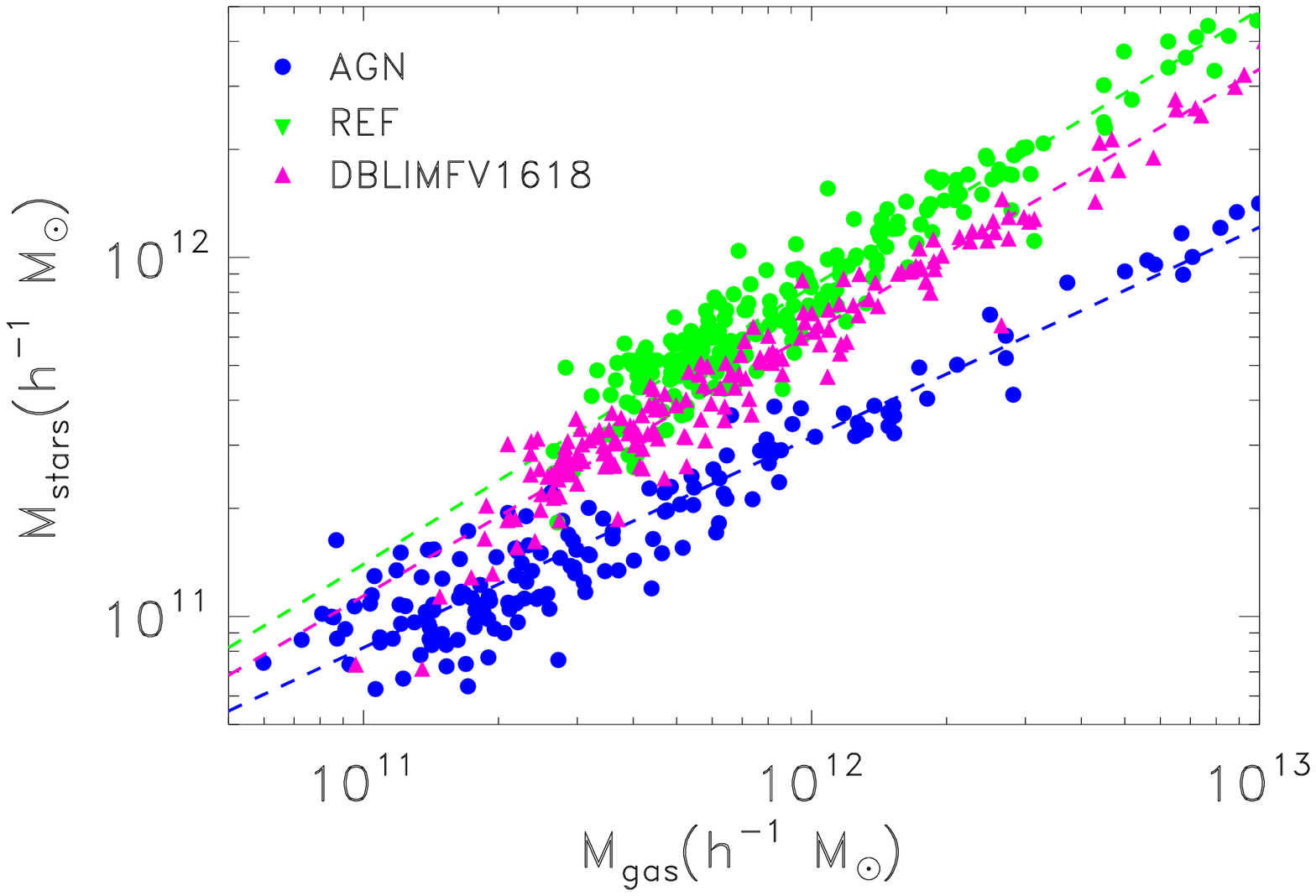,width=.50\textwidth}
\end{tabular}
\caption{\label{fig:fit}Left panel: relation between  the halo mass $M_{500}$ and the total baryonic mass contained within $r_{500}$ of the halo. The baryonic  mass content that corresponds to the universal baryonic fraction as a function of $M_{500}$  is indicated by the black solid line. 
 The dashed lines show the best fit power laws to the corresponding points. 
 The solid lines show the best fit to the points when we impose the constraint  that all the scenarios reach the universal baryonic fraction for a halo the same given mass $M_{\rm pivot}$ which is about $2.7 \times 10^{14} M_\odot/h$ ( more details about its derivation are given in the text). Note that for AGN and REF the solid and dashed lines are almost identical and therefore cannot be distinguished in the plot.
Right panel: relation between the hot gas and the stellar mass within $M_{500}$. The dashed lines represent the best-fit power laws to the data. }
\end{figure*}  

Throughout the  paper we use the halo model to predict the matter power spectrum analytically. The reason for this choice is that  the halo model framework allows one to naturally account  for the modifications to the power spectrum caused by  baryonic feedback.   
 As shown  by Seljak (2000) (see also Mandelbaum et al.\ 2005), the halo model reproduces the power spectrum into the non-linear regime quite well, although some parameters have to be calibrated
using numerical simulations. The power spectrum is computed as the sum of two terms: 
\be
P(k)=P^{\rm P}(k)+P^{\rm hh}(k)\,. 
\ee
\noindent The 
first term describes the correlation of the density fluctuations
within the same halo. This Poisson term $P^{\rm P}(k)$ dominates
on small scales and is given by

\begin{equation}\label{eq:one_halo}
P^{\rm P}(k)=\frac{1}{(2 \pi)^3} \int {\rm d}\nu f(\nu)
\frac{M(\nu)}{\bar\rho} y[k,M(\nu)]^2,
\end{equation}

\noindent where $\bar\rho$ is the mean matter density and $y[k,M(\nu)]$ is
the Fourier transform of the  density profile of a halo with
virial mass $M(\nu)$ which is normalised such that:

\be\label{eq:density_profile}
y[k,M]=\frac{\int_0^{r_{\rm vir}} 4 \pi r^2 {\rm d}r \frac{\sin(kr)}{kr} \rho(r)}{\int_0^{r_{\rm vir}} 4 \pi r^2 {\rm d} r \rho(r)}\,,
\ee

\noindent  where $\rho(r)$ is the density profile of haloes truncated at the virial radius $r_{\rm vir}$. The virial radius is defined  in practice as the radius within which the average density is greater than $\delta_{\rm vir}$ times the mean density of the Universe. We compute $\delta_{\rm vir}$ as in  Henry (2000) and Nakamura \& Suto (1997).
The peak height $\nu$ of such an overdensity is defined as
\begin{equation}
\nu=\left[\frac{\delta_{\rm c}(z)}{\sigma(M)}\right]^2,
\end{equation}

\noindent where $\delta_{\rm c}$ is the linear theory value of a spherical
overdensity  which  collapses at a redshift $z$ and $\sigma(M)$ is the rms of fluctuations in spheres that contain a mass $M$ at an initial
time,  extrapolated to $z$ using linear theory. To compute the mass function $f(\nu)$ we use \cite{ShTo99}:

\begin{equation}
\nu f(\nu)=A (1+\nu^{\prime-p}) \nu^{\prime 1/2} \exp({-\nu^\prime/2}),
\end{equation}

\noindent where $\nu^\prime=a\nu$ with $a=0.707$ and $p=0.3$. The
normalisation constant $A$ is determined by imposing $\int f(\nu) {\rm d}\nu
=1$. Note that the function $f(\nu)$ is related to the halo mass function ${\rm
  d}n/{\rm d}M$ through

\begin{equation}
\frac{{\rm d}n}{{\rm d}M}{\rm d}M=\frac{\bar\rho}{M}f(\nu){\rm d}\nu.
\end{equation}

\noindent The second term, $P^{\rm hh}(k)$, describes the clustering of haloes
and dominates on large scales. It is given by

\begin{equation}\label{eq:two_halo}
 P^{\rm hh}(k)=P_{\rm lin}(k) \Big ( \int {\rm d}\nu f(\nu) b(\nu) y [k,M(\nu)]
\Big)^2,
\end{equation} 

\noindent where $P_{\rm lin}(k)$ is the linear power spectrum, and
the halo bias $b(\nu)$ is given by \cite{Maetal05}:

\begin{equation}
b(\nu)=1+\frac{\nu^\prime-1}{\delta_c}+\frac{2p}{\delta_c(1+\nu^{\prime
    p})},
\end{equation}

\noindent with $a=0.73$, $p=0.15$ and  $\nu^\prime=a \nu$.  To predict the weak lensing signal one needs to integrate Equation (\ref{eq:one_halo}) and (\ref{eq:two_halo}) in a range of masses that is wide enough. We take $10^8 <M_{\rm vir}< 10^{16} M_\odot/h $. Larger haloes have a very small probability to form whereas small haloes do not contribute significantly to the weak lensing signal.
 The last ingredient that we have not yet specified is the density profile of the matter haloes. Numerical cold dark matter
simulations have shown that the NFW profile \cite{Naetal95}

\begin{equation}
\rho(r)\propto\frac{1}{r(r+r_s)^2},
\end{equation}

\noindent where $r_s$ is the scale radius, is a fair description of
the radial matter distribution for haloes with a wide range in mass.
They also indicate that $r_s$ is not a free parameter, but that it is
related to the virial mass (albeit with considerable scatter). It is
customary to account for this correlation by specifying a relation
between the concentration $c=r_{\rm vir}/r_s$ and the virial mass. 
When baryonic physics is included, the dark matter profiles remain fairly well  approximated   by  NFW profiles (see for example Duffy et al.\ 2010), although the mass-concentration relation  changes due to the back-reaction of the baryons on the dark-matter component.
%  The virial mass and radius are related through 
%$M_{\rm vir}=(4\pi/3)r_{\rm vir}^3{\rho_c}\delta_{\rm vir}$, where we use the fittingformula of ... to compute $\delta_{\rm vir}(z)$.
\begin{figure*}
\begin{tabular}{|@{}l@{}|@{}l@{}|} 
\psfig{figure=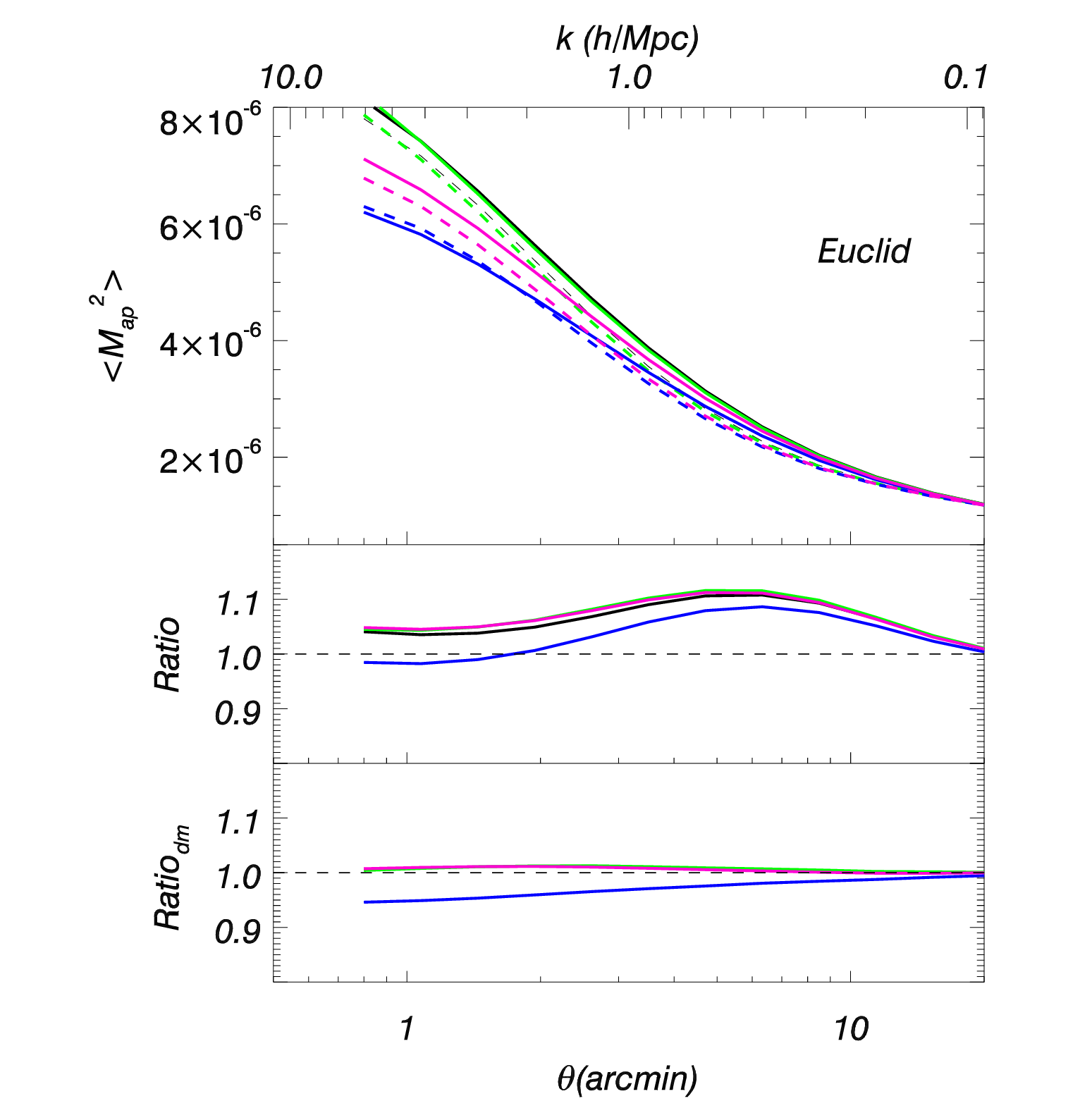,width=.50\textwidth}&\psfig{figure=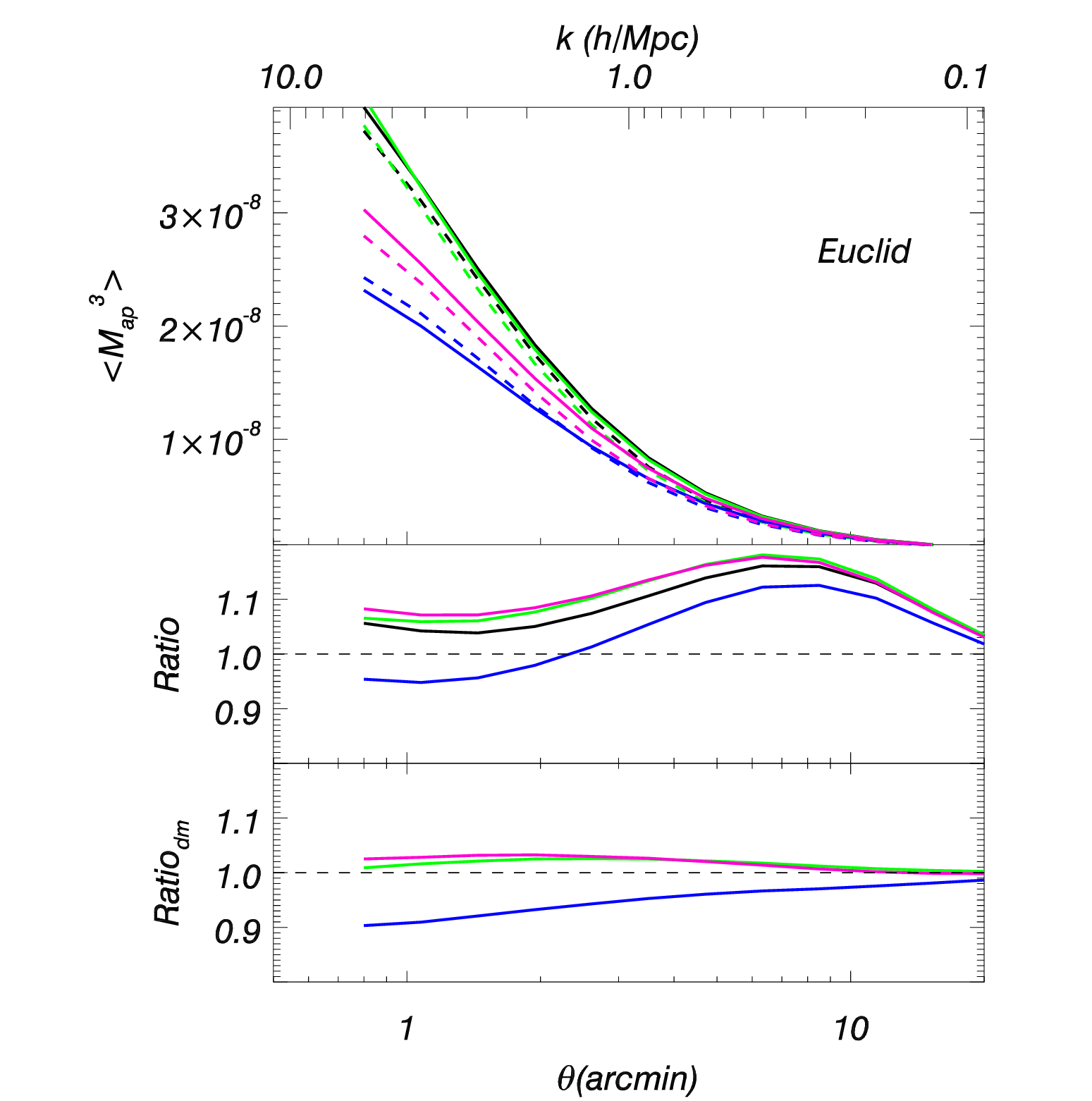,width=.50\textwidth}
\end{tabular}
\caption{\label{fig:map_hod} The top-left panel shows the amplitude of  $\ave{M_{\rm ap}^2(\theta)}$ as a function of the angular scale $\theta$ for the DMONLY (black), REF (green), DBLIMFV1618 (pink) and AGN (blue) simulations compared with the amplitude predicted using our modified halo model (dashed lines).  The top-right panel shows the same results as the left panel but for $\ave{M_{\rm ap}^3(\theta)}$. The middle panels show explicitly the ratio between  the value measured in the simulations and the predictions from the modified halo model and the bottom panels show the same ratio after the common  inaccuracy of the halo model  has been accounted for (see text). We indicate on the top the comoving wave number $k$ that  contributes most to the values of  $\ave{M_{\rm ap}^2(\theta)}$  and $\ave{M_{\rm ap}^3(\theta)}$ assuming that the sources are at redshift   $z=1$.  The plots show the good agreement between the cosmic shear signal measured for any of the scenarios and the predictions from the extended halo model developed in this paper. In particular, the middle and bottom panels show that this model  describes  the cosmic shear signal in any of the feedback scenarios with an accuracy  that is comparable to the accuracy with which the traditional  halo model describes the signal in a dark matter only scenario.}
\end{figure*}  

We have already shown in Semboloni et al.\ (2011a) that  modifying the traditional NFW profile in the halo model to account for gas profiles and star fractions provides a good description of the power spectrum predicted in each feedback scenario. A somewhat different approach has been developed by Zentner et al. (2008, 2012) who account for  the effect of feedback by changing the mass-concentration relation.  The advantage of the latter approach is that halo profiles are still parametrised with only two parameters although the mass concentration relation changes as a function of redshift. However, if the hot gas and the dark  matter have a very different distribution, then this approximation might not be accurate. Furthermore,  it is more difficult to constrain using observations than the model  developed by Semboloni et al.\ (2011a) that we employ  here.

We assume that each halo is collapsing with an initial fraction of baryons which is equal to the universal gas fraction, $f_{\rm gas}=\Omega_b/\Omega_m$.  We compare this initial fraction to the fraction of baryons (hot gas plus stars) contained in $r_{500}$, defined as the radius of a sphere within which the mean density equals  500 times the critical  density.  We select the same haloes as Semboloni et al.\ (2011a) which are also used in McCarthy et al. (2010). These haloes cover a mass range $5 \times 10^{12} M_{\odot}/h < M_{500} < 10^{14}M_{\odot}/h$. This covers the regime of halo masses  to which cosmic shear is most sensitive (see for example Semboloni et al. 2011a). In this range,  the relation between the total mass of baryons contained within $r_{500}$  versus the total halo mass $M_{500}$  is fairly well described by a power-law  (see left panel of Figure \ref{fig:fit}): 
\be \label{eq:linear}
\log_{10}(M_{\rm bar})=a_{\rm bar}\log_{10}(M_{500})+b_{\rm bar}\,.
\ee
\noindent  A power law can also be used to describe the relation between the hot gas (i.e. gas with temperature $T> 2.2 \times 10^{6} K$ ) and the stellar mass contained in haloes  (see right panel of Figure \ref{fig:fit}):
\be \label{eq:stars}
\log_{10}(M_{\rm star})=a_{\rm star}\log_{10}(M_{\rm gas})+b_{\rm star}\,.
\ee
For each scenario we use these relations to determine the stellar and hot gas fractions contained in haloes.  We fit these functions to the results from McCarthy et al. (2010, 2011). We extrapolate  the results  for lower and higher masses which are needed for the halo model.  Although this extrapolation at low masses might not be accurate,  it will affect the weak lensing predictions only marginally as small masses  ($M_{500} < 10^{12} M_\odot/h$)  do not contribute significantly to the cosmic shear signal. When we extrapolate to high masses  we reach a mass for which the  baryonic fraction is equal to the universal value. We know from the observations that for higher masses  the gas fraction does not exceed   the universal fraction and we therefore fix it to this value.

 We describe the spatial distribution of each of the three components in the halo using different profiles. We approximate the stellar profile by a point mass. For the dark matter component we use an NFW profile. The hot gas is described by a $\beta$-model \cite{CaFu76} 
\be
\rho_{\rm gas}=\rho_0\Big[1+\Big(\frac{r}{\alpha r_{500}}\Big)^2\Big]^{-3\beta/2}
\ee
 where the parameter $\alpha$ describes the relation between $r_{500}$ and the characteristic scale of the  gas profile, and $\beta$ is the characteristic slope of the profile.   The total mass of the gas within $r_{500}$ determines the value of $\rho_0$. The values of $\alpha$ and $\beta$ are measured from the simulations.
The gas that has been moved beyond the virial radius is distributed uniformly in a region between $r_{\rm vir}<r<2 r_{\rm vir}$.  This choice is partially motivated by  independent studies that suggest the presence  of  gas in the outskirts of haloes, but it is mostly a practical choice allowing us to modify halo profiles  without inserting a new component. The validity of this assumption should be investigated further, but it does not have a major impact on the conclusions of this paper. 

The final profile is then the sum of the individual components:
\be
y_{\rm tot}[k,M]=\sum_{i=1}^n y_i[k,M] \frac{M_i}{M}
\ee
\noindent where $i$ identifies the component of the halo profile, which can be  dark matter, hot gas  within the viral radius, stars  or the smooth gas component  beyond the virial radius. The sum of the masses of the four components is the virial mass. The profiles $y_i[k,M]$ of the dark matter and hot gas components  are obtained by truncating the density profiles at the virial radius as in Equation (\ref{eq:density_profile}). The expression  $y_i[k,M]$ for the smooth gas component beyond the viral radius is obtained by changing  the integration range in Equation (\ref{eq:density_profile}) to $r_{\rm vir}<r<2 r_{\rm vir}$.

To predict the DMONLY model, we use the halo model with the usual NFW profile and the mass-concentration relations derived by Duffy et al.\ (2008). Note that Duffy et al.\ (2008) used a  WMAP5 cosmology which is different form the OWLS WMAP3 cosmology. Since the mass-concentration relation depends on the cosmology (on $\sigma_8$ in particular, e.g. Eke et al.\ 2001) this can introduce a difference between the measured and predicted spectra. In the case of baryonic feedback we use the  mass-concentration modifications  derived  by Duffy et al.\ (2010) using the OWLS models in the following way: the  differences between feedback models and the dark matter only scenario as a function of redshift are obtained by linear interpolation of  the average of the ratios shown in that paper at redshifts $z=0$ and $z=2$.  For  DBLIMFV1618  Duffy et al.\ (2010) do not provide the mass-concentration relations and therefore we chose to use the ones that are derived for a strong SN feedback model (``ZC\_SFB'' in Duffy at al.\ 2010) which is denoted as WDENS here.  
The $\alpha$ and $\beta$ parameters are also fixed to the average value measured in the simulations. For  DBLIMFV1618, for which we could not measure these parameters directly,  we fix these values  to the  AGN scenario as we found that this provides us with a better model than when using the REF profile.

In Figure \ref{fig:map_hod} we show the signal measured in the simulations  compared with the predictions obtained using  the halo model,  both for $\ave{M_{\rm ap}^2(\theta)}$  and $\ave{M_{\rm ap}^3(\theta)}$  for a survey like {\it Euclid}. Since we are using the wrong mass-concentration relations and we are  fixing the stellar fraction, the gas fraction and the gas profiles at $z=0$,  we do not expect our model to work perfectly. Nevertheless,   the model can reproduce the general features for all feedback scenarios quite well. In particular, our approach reproduces $\ave{M_{\rm ap}^2(\theta)}$ and $\ave{M_{\rm ap}^3(\theta)}$ predicted by the simulations to within $10\%$ and $20\%$  respectively, for any of the baryonic feedback descriptions. This accuracy is similar to the one with which the dark matter only halo model  describes the DMONLY  results. This is shown more explicitly in the middle-left panel where we plot the ratio
\be
R\equiv\frac{\ave{M_{\rm ap}^2(\theta)^{\rm OWLS}}}{\ave{M_{\rm ap}^2(\theta)^{\rm hod}}}\,,
\ee 
i.e. the ratio of the values measured in the simulations and the ones predicted by the extended halo model. In the right-middle panel, we show the same results for  $\ave{M_{\rm ap}^3(\theta)}$. 

The lack of accuracy of our model at intermediate scales,  about $10\%$ for $\ave{M_{\rm ap}^2(\theta)}$ and $20\%$ for $\ave{M_{\rm ap}^3(\theta)}$ at scales of $\theta \sim 10~{\rm arcmin}$, is mainly caused by  the  lack of agreement between the halo model and the DMONLY results.  The limited accuracy with which  the halo model  in its original formulation  predicts  dark matter  power spectra has  been discussed in a number of papers (see for example Hilbert et al. 2009) and it is beyond the scope of this paper to improve the model or to discuss its performance. It is, however, clear that its performance needs to be improved before it can be used to interpret data from large weak lensing surveys. The ``halo-exclusion'' \cite{Tietal05,Caetal09}  and  in particular the perturbative halo model approach \cite{VaNi11} are notable steps in this direction. 
 Note, however, that the discrepancy  between  the  DMONLY and halo model prediction is not necessarily due  only to the inaccuracy of the halo model.  The DMONLY results are affected by cosmic variance (see appendix A1 of van Daalen et al.\ 2011). Furthermore, as we have  mentioned already, the mass-concentration relations we have used are not the most appropriate for the WMAP3 cosmology assumed in the simulations.
 
To avoid any issues originating from the direct comparison between DMONLY and halo model predictions, we will follow Semboloni et al. (2011a) and work only with ratios. The advantages of using ratios is that we can suppress the effect of cosmic variance and finite size  but explore the effect of the baryonic feedback up to small $k$.   For this reason we divide for each scenario the ratio $R$ by the same ratio obtained for the DMONLY scenario  and show these results in the bottom panels.
In this way, one can see better  how well the model captures the modifications introduced by our feedback model. The corrections we have applied to the halo model  reproduce the weak lensing signal from the REF and DBLIMFV1618 scenarios very well, whereas  the effect of baryonic feedback for the AGN scenario is slightly underestimated. However, the difference is only a few percent for  $\ave{M_{\rm ap}^2(\theta)}$  and ten percent in the worst case for  $\ave{M_{\rm ap}^3(\theta)}$.

\subsection{A general description of feedback}
The model we have described in the previous subsection has enough flexibility  to  adequately describe  any of the OWLS scenarios we have examined. In this model, we have assumed that the effect of baryonic feedback  can be captured by six parameters: two  describing the shape of the gas profile ($\alpha$ and $\beta$), two  describing the total mass  of  baryons  (stars + hot gas) contained in a halo of mass $M_{500}$, and two  describing the stellar mass as a function  of the mass of the hot gas. Furthermore, for each simulation we used  the mass-concentration relations  from Duffy et al.\ (2010).
Since the real strength of the feedback is not known, one could imagine  marginalising over these parameters to obtain unbiased cosmological constraints. However, one would like to keep the number of nuisance parameters as small as possible, both for computational and signal-to-noise purposes.
The aim of this section is to check whether it is possible to reduce the number of degrees of freedom in this model and to see if in doing so we are able to describe the impact of feedback generically.

\begin{figure}
\psfig{figure=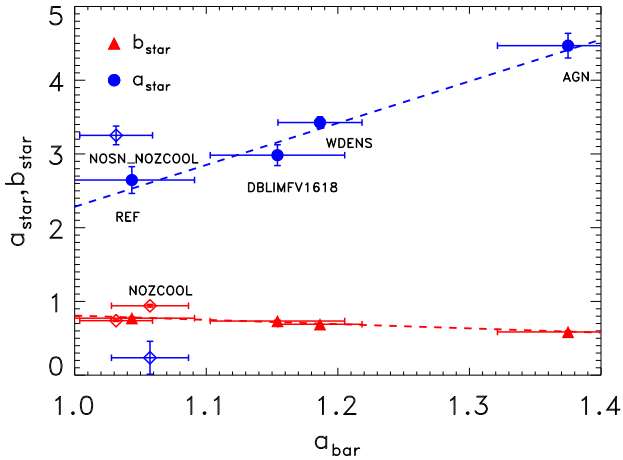,width=.50\textwidth}
\caption{\label{fig:relations} Blue solid circles  (solid red triangles) show the value of  $a_{\rm star}$ ($b_{\rm star}$) describing the fraction of stars in each halo (see Equation \ref{eq:stars}), as a function  of the total baryonic content  expressed by  the value of  $a_{\rm bar}$  (see Equation \ref{eq:fgas}). From left to right the values of $a_{\rm bar}$ correspond to REF, DBLIMFV1618, WDENS  and AGN. The error bars  provide the errors from each individual fit.   Note that if $a_{\rm bar}=1$ the baryonic fraction is the same as the universal gas fraction for all haloes.   The dashed lines are the best linear fits to the data from  REF, DBLIMFV1618 and AGN.  These fits can capture the general effect of feedback on the stars and hot gas distribution in the haloes also for the WDENS simulation.  Open diamonds indicate the results for the NOZCOOL and NOSN\_NOZCOOL scenarios, which do not follow the trend displayed by the other models.}
\end{figure}

As a first step, we show below that the fraction of stars and the fraction of hot gas as a function of halo mass are correlated. In fact, looking at Figure \ref{fig:fit},  one can see that for any given feedback model there is a relation between halo mass, mass of hot gas and mass of stars; if the feedback is stronger there is less gas and there are also fewer stars. This suggests that for a given set of parameters describing the gas fraction there may be a set of parameters  which defines  the stellar mass.
Furthermore, the halo mass for which the fraction of baryons inside the halo equals the universal fraction is very similar for all scenarios. The new fit performed, fixing the `pivot' mass based on the AGN model ($\log_{10}(M_{\rm pivot}/M_\odot)=14.43$), is shown in the left panel of  Figure \ref{fig:fit}  (solid lines) and can be compared with the original fit performed for each scenario individually (dashed lines).  The fit for the REF scenario is unchanged, while  the DBLIMFV1618 is slightly tilted, but the quality of the fit is the same and the differences are minor. Replacing the expression of $M_{\rm pivot}$ in Equation (\ref{eq:linear}) we obtain:
{\setlength\arraycolsep{0.1em}
\ba\label{eq:fgas}
\log_{10}(M_{\rm bar})&=&a_{\rm bar} [\log_{10}(M_{500}) -\log_{10}(M_{\rm pivot})]\\
&+&\log_{10}(M_{\rm pivot})+\log_{10}(f_{\rm gas})\nonumber
\ea
}
which assigns to any halo of mass $M_{500}$ a universal gas fraction if $a_{\rm bar}=1$, and a lower baryonic mass for a given halo of mass $M_{500}$ as the slope $a_{\rm bar}$ increases. As before, for masses larger than $M_{\rm pivot}$ we keep the gas fraction fixed to the universal value. 
 We cannot reduce the number of parameters for the stellar fraction, which we continue to parametrise with $a_{\rm star}$ and $b_{\rm star}$. However, Figure \ref{fig:relations} shows  that there is a linear relation between the slope $a_{\rm bar}$  and the parameters  $b_{\rm star}$ and $a_{\rm star}$. That suggests that independently of the mechanisms that remove the gas, there is a tight relation between star and gas content provided that a single process is responsible for expelling the gas and suppressing the star formation.
Figure \ref{fig:relations} also suggests  that this relation can be expressed in a rather simple way. In order to check  the validity of our relation,  we also show the results of the best-fit obtained for the  WDENS simulation. Although  the linear  relations  in Figure \ref{fig:relations} have been obtained using only the simulations REF, AGN and DBLIMFV1618,  the WDENS best-fit values  seem to agree well with this relation.

As we  anticipated, the WDENS simulation  has a power spectrum modification which is very similar to the one of DBLIMFV1618 (see van Daalen et al.\ 2011).
However, the physical mechanisms  responsible for changing the large-scale matter distribution  are different and it is remarkable  that  even in this case the gas and stellar content seem to follow the same law that we found for the other simulations. This suggests that there might be some physical reason behind our findings. 

The correlation between the parameters breaks down for the NOZCOOL and NOSN\_NOZCOOL simulations. 
 Whereas it is true that simulations which do not implement metal enrichment and metal-line cooling and simulations without any energetic feedback from star formation are unrealistic, it is not necessarily true that including these processes allows one to obtain simulations with   realistic gas properties (the REF simulation is an example).  However, in the simulations where metal-line cooling  and supernova feedback are included  the conditions  for  star formation  are  similar and  the injection of energy in the gas followed by its displacement determines how frequently these conditions are met. This common mechanism seems to be responsible for the relations we have found. Whether these simple relations are able to describe a large class of realistic feedback scenarios is a question that needs to be addressed with  other sets of simulations using a wider variety of feedback implementations and is beyond the scope of this paper.

If these relations were proven to be generic, they could  provide us with a recipe to  determine the  hot-gas and the stellar fractions within a halo of a given mass.   This would mean that the feedback could be described by a single  parameter: the fraction of gas left in a halo as a function of its mass. The parameters of the $\beta$-model and mass-concentration relation characterise the overall distribution of matter within  the haloes and also change the predicted power spectrum. In principle, they also  depend on the specific feedback model, but we expect them to be fairly degenerate. For this reason, we keep $\alpha$ and $\beta$ as free parameters in our model when we perform a likelihood analysis in the next section.  The ranges are chosen to include values  measured in the  simulations and are $\alpha=[0.0005,0.15]$ and $\beta=[0.45,0.85]$. We do  fix  the mass-concentration parameters to those for the AGN simulation. This is just one possible  approach to marginalise over the lack of knowledge of the overall profile. One could also fix the $\beta$ profile and change the NFW parameters or vary them both within the ranges set by the simulations. Note that the NFW parameters do not change much between  the various feedback scenarios,  while the $\beta$ profile parameters do.  The dispersion in those parameters is also large and depends on the mass. However,  since the hot gas fraction is small, the modifications to the overall profile have only a minor impact on the power spectrum: to first order all the modifications  arise because of the removal of the hot gas. 

\subsection{Cosmology dependence}
\begin{figure}
\psfig{figure=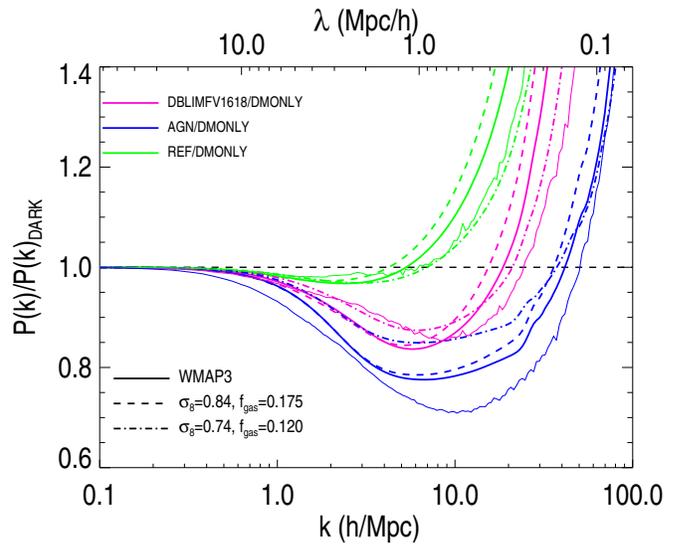,width=.50\textwidth}
\caption{\label{fig:model_cosmo} Ratio between the dark matter power spectra predicted by our halo model for various baryonic scenarios and the dark matter only case,  for $z=0$ and different cosmologies.  The WMAP3 cosmology results (solid lines) are compared to the same results obtained for a WMAP3 cosmology but with $\sigma_8=0.84$ (dashed lines). In this case $f_{\rm gas}$ is the same as the WMAP3 cosmology and the differences are small. The dot-dashed lines show the results for a WMAP3  cosmology  but with  $f_{\rm gas}=0.120$. In this case the ratios in the case of the DBLIMFV1618 and AGN scenarios are significantly changed. The solid thin lines represent the ratio measured in the simulations (which use the WMAP3 cosmology) to show the accuracy of the model used in this paper.}
\end{figure} 

Finally, we investigate the cosmology dependence of the feedback model adopted in this paper.  To do so we compute the  ratio of power spectra $P(k,z)/P^{\rm DM}(k,z)$ predicted by our halo model   for different cosmologies.  Whereas, the universal gas fraction $f_{\rm gas}$ fixes  the fraction of gas  present in each halo at the moment of collapse, Equation  (\ref{eq:fgas}) determines which fraction of this initial mass is still within the halo at low redshift and which fraction is ejected beyond the virial radius.  For this reason any change in $\Omega_m$ leads to the same modification of the power spectrum as long as $f_{\rm gas}=const$.  Note that, since we do not know how the profile parameters vary as a function of cosmology, we keep them fixed. 
 
When we change the normalisation of the matter power spectrum $\sigma_8$, our halo model predicts that the ratio of the power spectra  changes slightly. This differs from   van Daalen et al.\ (2011) who have shown that, in the case of the AGN scenario,  the relative variation to the dark matter only power spectrum is  very close for the WMAP3 and WMAP7 cosmologies. We remind the reader that these cosmologies  differ  mainly because of the  change of  $\sigma_8$.  The discrepancy between our results and the ones in van Daalen et al. (2011) can be understood as follows.
Changing the value of $\sigma_8$ changes the halo mass function. One can see from Equations (\ref{eq:one_halo}) and (\ref{eq:two_halo}) that the halo mass function  acts as a weight  applied to any profile of mass $M$. The ratio between the total power spectra is given by the ratio of the weighted sums.
Since we keep the profile  parameters unchanged when we change the cosmology, it is not surprising that  the ratio varies.  On the other hand, we know that the profile characteristics vary with the cosmology. In particular, smaller values of $\sigma_8$ lead to the formation of haloes with a smaller concentration, at least in DMONLY scenarios. 
It remains to be seen if our model is able to reproduce the result  by van Daalen et al.\ (2011) when the profiles are assigned correctly as a function of cosmology, something which we avoid doing here because we do not know how the profiles vary as a function of cosmology, especially when  baryonic feedback is included.

We show in Figure \ref{fig:model_cosmo} the ratio of the power spectrum  in each scenario to the dark matter-only scenario for two different cosmologies, and compare these with the WMAP3 results. As one can see  by comparing the solid lines (WMAP3 cosmology) and dashed lines (WMAP3 cosmology with different $\sigma_8$ value), varying $\sigma_8$ while keeping the universal baryonic fraction constant leads only to small variations of the ratio.   When the universal gas fraction is changed (dot-dashed lines in Figure \ref{fig:model_cosmo}) the modifications to the power spectrum are also changed. The variation is not large in the case of the REF scenario as the fraction of gas that leaves the haloes is small, but it is larger in the case of the AGN and DBLIMFV1618 scenarios. In particular, when the universal gas fraction decreases, the effect of the feedback is reduced as  the  dark matter fraction in the halo is increased.

Finally, we note that this model is different from the one we used in Semboloni et al.\ (2011a) because we  use different scaling relations. In particular, we now fit the stellar masses of the halos  directly, whereas Semboloni et al.\ (2011a) fitted the luminosity and used observed mass-to-light ratios to derive the stellar fraction. Because of the changes and the dispersion in the fitted relations, which for simplicity were ignored in both papers, one expects the model to perform slightly differently, although this does not change the conclusions of Semboloni et al.\ (2011a). The description adopted here is  more general than the one adopted in Semboloni et al.\ (2011a) as Equation (\ref{eq:fgas}) provides the general relation needed to modify the dark matter power spectrum in the way we have just shown.

\begin{figure*}
\begin{tabular}{|@{}l@{}|@{}l@{}||@{}l@{}|} 
\psfig{figure=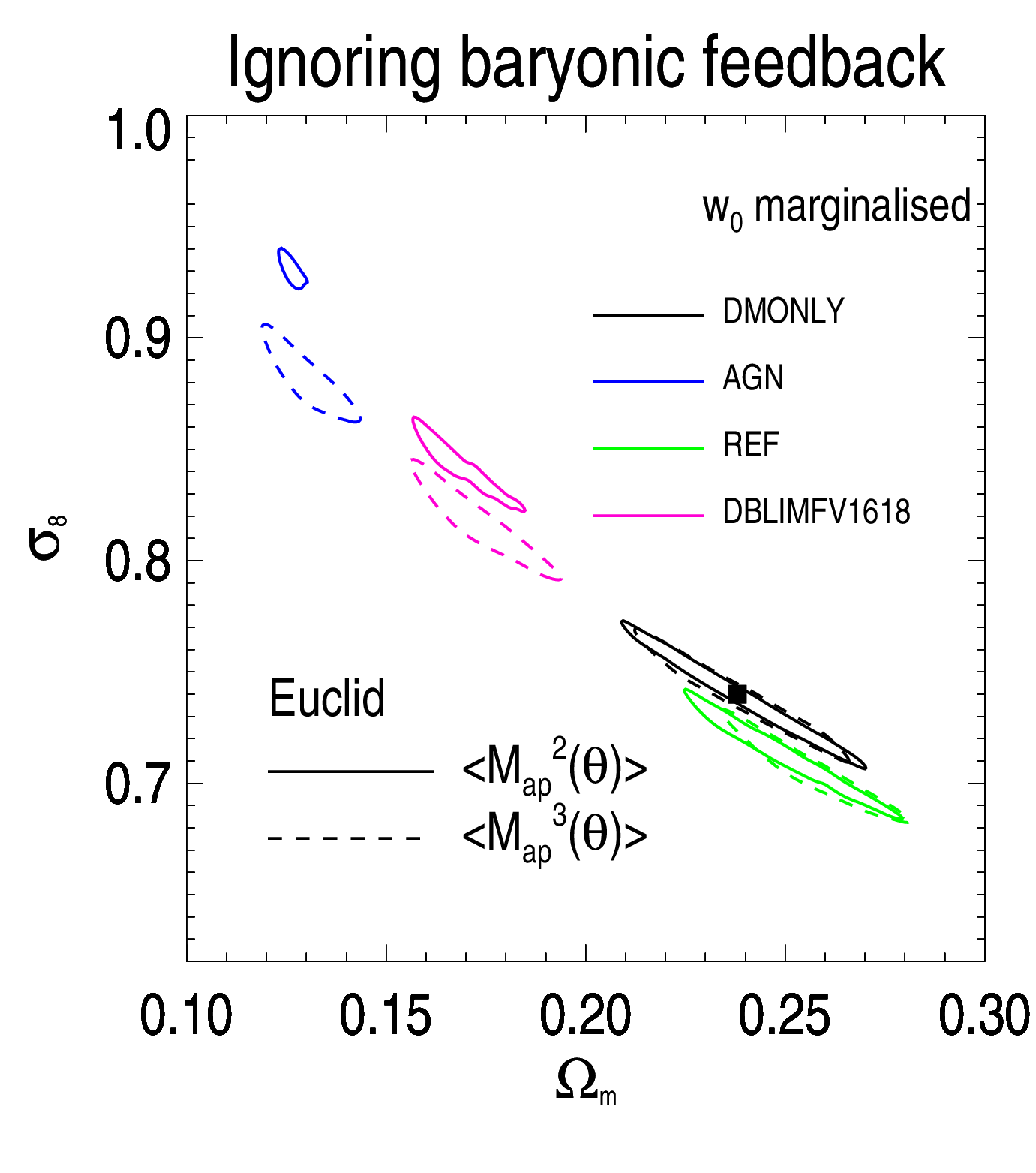,width=0.33\textwidth}&\psfig{figure=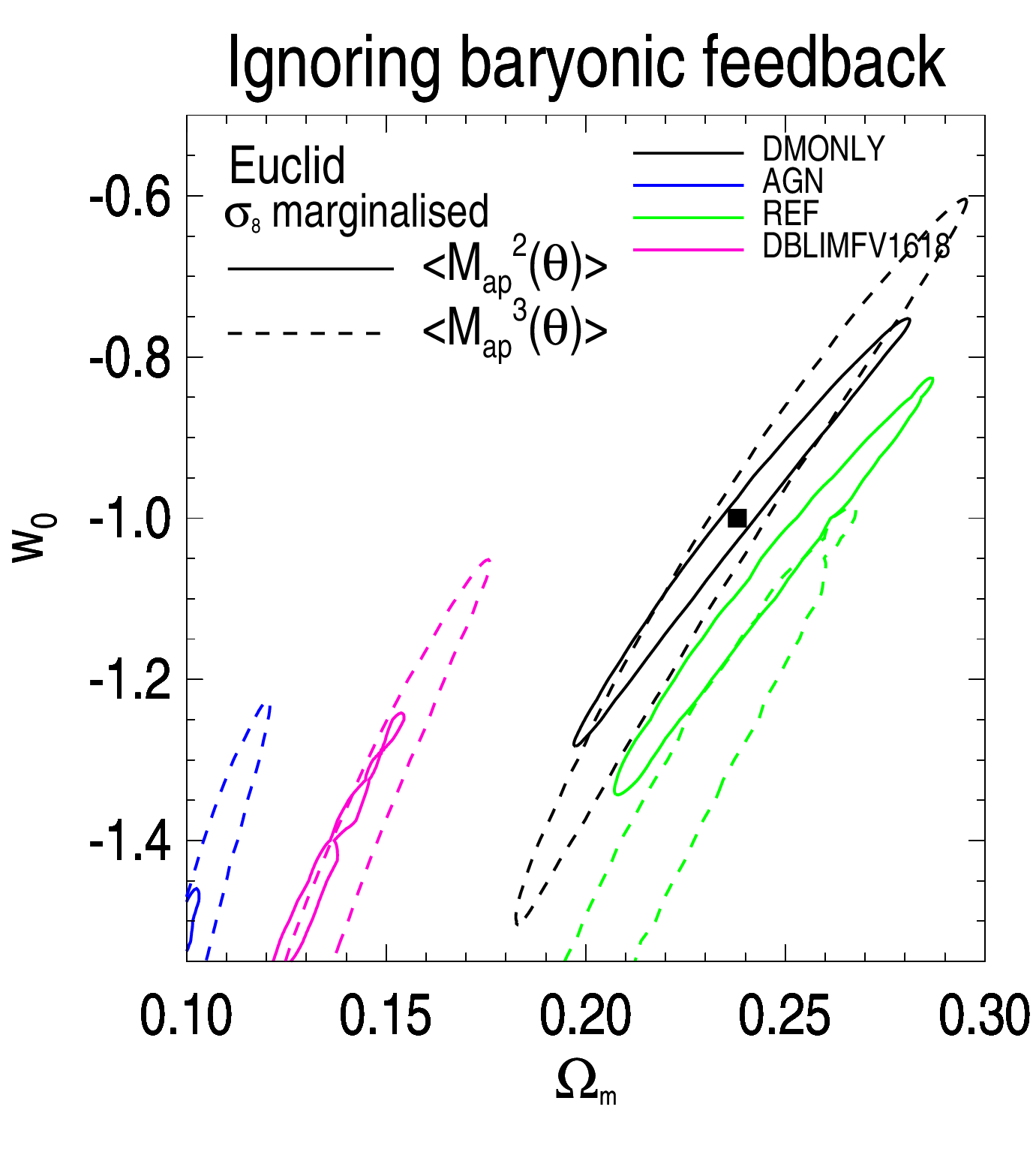,width=0.33\textwidth}
&\psfig{figure=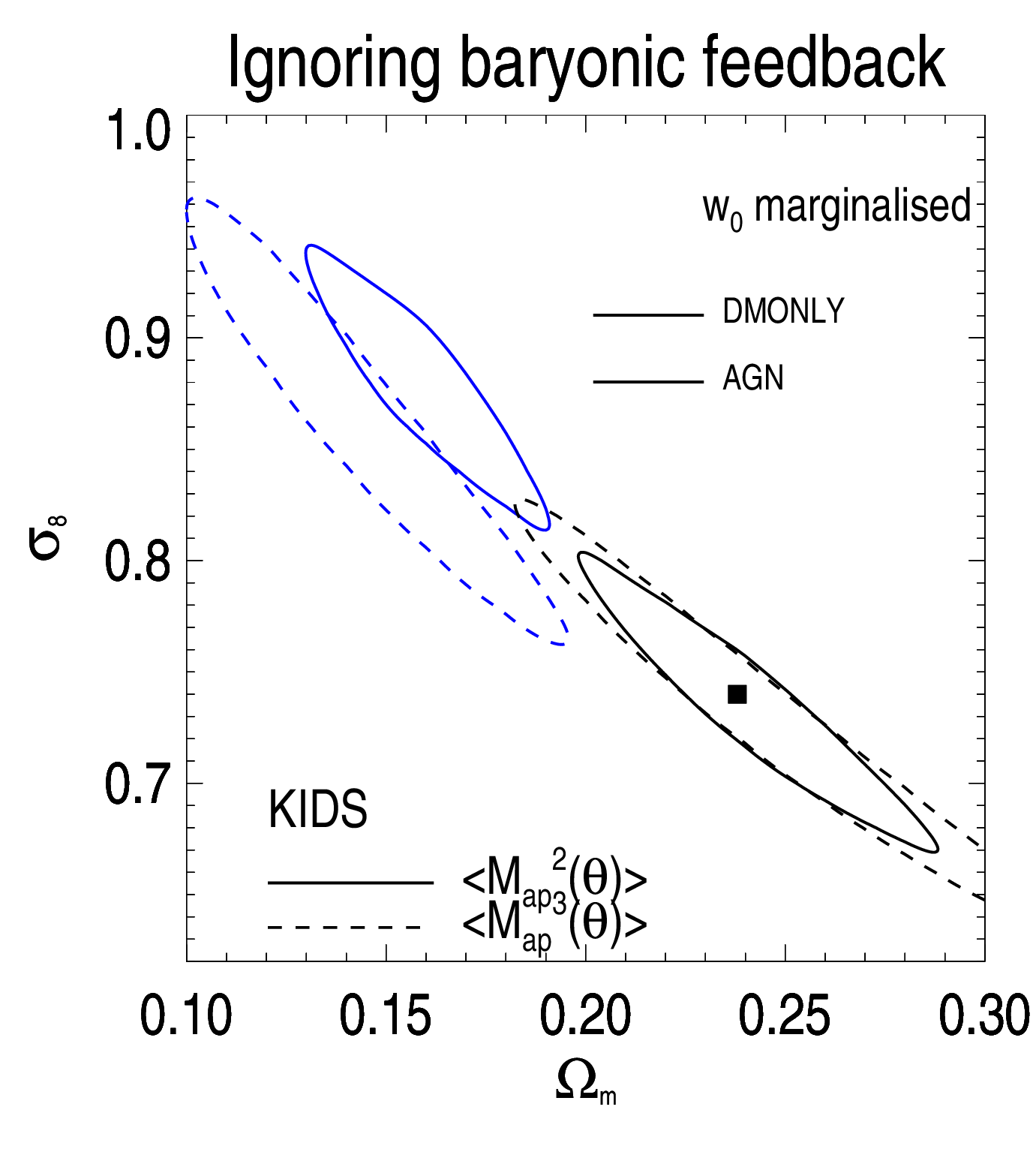,width=0.33\textwidth}
\end{tabular}
\caption{\label{fig:fisher} Left panel:  likelihood analysis results for $\ave{M_{\rm ap}^2(\theta)}$ (solid lines) and $\ave{M_{\rm ap}^3(\theta)}$ (dashed lines) for the DMONLY (black), REF (green), DBLIMFV1618 (pink)  and AGN (blue) scenarios. The $1\sigma$ contours  have been obtained for the {\it Euclid}-like fiducial survey. The square indicates the WMAP3 cosmology which should be the maximum of the probability distribution in case of unbiased results. Middle panel: likelihood analysis results in the $\Omega_m$-$w_0$ parameter space obtained for the {\it Euclid}-like fiducial survey.
Right panel: same as the left panel but for a KiDS-like survey. For the KiDS-like survey we show only the deviations for the DMONLY and AGN scenarios.  The models do not include baryonic feedback and have been computed using the halo model power spectrum and rescaled as described in the text. Any shift relative to the DMONLY case indicates the presence of bias due to baryonic effects.}
\end{figure*}  

\section{Likelihood analysis}\label{sec:likelihood}

 In this section we will show that the combined measurement of two- and three-point shear statistics on large datasets can unveil the existence of  strong baryonic feedback. 
 Furthermore, we will  show that the  effect of an unknown feedback scenario can be mitigated, provided that it can be modelled as described in the previous section.
   To do that, we  perform a likelihood analysis  where the cosmological parameters $\Omega_m$, $\sigma_8$ and $w_0$ are varied. All other parameters are fixed to their value for a WMAP3 cosmology, while we set $w_a$  to zero. The posterior probability is given by:  
\be
P({\bf \pi} | {\bf d}) \propto  \mathcal{L} ( {\bf d} | {\bf \pi} ) P({\bf \pi})\,, 
\ee
where:
\be
{\mathcal L }( {\bf d} | {\bf \pi} )=\exp [{-1/2 ( {\bf d}- {\bf m}({\bf \pi}))^t   \mathbfss{C}^{-1} ({\bf d}- {\bf m}({\bf \pi}))}]
\ee
and we have indicated with ${\bf \pi}$, ${\bf d}$ and ${\bf m}$ the parameter space, the data  and the model vectors, respectively.  

 To compute the covariance matrix $\mathbfss{C}$, we use the set  of projected density  maps  described in Vafaei et al.\ (2010) which have been realised with the same WMAP3 cosmology. Each convergence map consists of  $1024^2$ pixels covering a total area of  $12.84~ {\rm deg^2}$. We have  $60$ lines-of-sight and along each line-of-sight the projected density field is sampled in $40$ steps between $0<z<3$. For any line-of-sight,  we  combine the $40$ projected convergence maps using the redshift distribution of each survey we are interested in. Finally, we add shot noise to the $60$ convergence maps assuming  a galaxy density of $10~ {\rm galaxies/arcmin^2}$  for the KiDS-like and $30~{\rm galaxies/arcmin^2}$ for the {\it Euclid}-like   survey.  The intrinsic ellipticity dispersion is set to $0.2$ per component. We use an area $A=15000\, {\rm deg}^2$  for  the {\it Euclid}-like survey and $A=1500\, {\rm deg}^2$ for KiDS.   We use  $12$ angular scales in the range  $0.5$ to $20$ arcmin. The maximum angular scale is then smaller than the side of each map and  we can rescale the covariance matrix by multiplying  by the ratio of the area of the  maps   to the area of the survey we want to simulate. In fact, for a given survey depth and galaxy density, the overall statistical error scales $\propto A^{-1/2}$, assuming that boundary effects can be neglected and that the survey is composed of contiguous observations (see for example Schneider et al.\ 2002). Strictly speaking, this is only true in the Gaussian regime, and since we apply this simple rescaling our covariance matrix is understimated (see Sato et al. 2009 and Kayo et al. 2013).  Although it is important for the future  to either simulate very large volumes  or to use analytic formula as suggested by Kayo et al. (2013), this is not a major issue for this paper.

To compute the	 predicted cosmic shear signal,  we use the  linear power spectrum of density fluctuations derived with  the transfer function from Eisenstein \& Hu (1998). The non-linear evolution is derived using our halo model. The data vector ${\bf d}(\theta)$  is derived by computing $\ave{M_{\rm ap}^2(\theta)}$  and  $\ave{M_{\rm ap}^3(\theta)}$  for each  OWLS scenario. To alleviate effects arising from the lack of agreement between halo model (dark matter-only) predictions and the DMONLY simulation, we follow Semboloni et al.\ (2011a) and  we recalibrate the data vector ${\bf d(\theta)}$:
\be
{\bf d}(\theta)= \frac{{\bf d}(\theta)^{\rm hod}}{{\bf d}(\theta)^{\rm DMONLY}} {\bf d}(\theta)^{\rm OWLS}
\ee
where ${\bf d}(\theta)^{\rm OWLS}$ is the data vector measured in the baryonic OWLS simulations and ${\bf d}(\theta)^{\rm DMONLY}$ is the data vector measured for the dark matter only simulation. This ensures that  when the input vector is the one measured in the DMONLY scenario, the data vector we use in the likelihood analysis is the one obtained using ${\bf d}(\theta)^{\rm hod}$, the halo model with the same cosmology as OWLS.

 We show in the left panel of  Figure \ref{fig:fisher} the  $1\sigma$  error contours in  $\Omega_m$-$\sigma_8$ parameter space inferred from the measurement of $\ave{M_{\rm ap}^2(\theta)}$ (solid contours)   and  $\ave{M_{\rm ap}^3(\theta)}$ (dashed contours)  for the {\it Euclid}-like survey.  They have been obtained after marginalisation over a flat prior on $w_0=[-1.2,-0.8]$ which we assume can be inferred  from non weak-lensing experiments. For this same fiducial survey we show in the middle panel of Figure \ref{fig:fisher}, the constraints in the $\Omega_m$-$w_0$ parameter space after marginalising over $\sigma_8$.
  The left and middle panels of Figure \ref{fig:fisher} show that the interpretation of both statistics  from the {\it Euclid}-like survey, using a dark matter only model, provides a probability  distribution of the posterior which is increasingly biased when the  feedback becomes stronger.  Furthermore, there is a discrepancy  between the posterior probability distributions  for $\ave{M_{\rm ap}^2(\theta)}$  and  $\ave{M_{\rm ap}^3(\theta)}$ which is more evident in the case of $\Omega_m$-$\sigma_8$ constraints. The consequence of  baryonic feedback is to displace the posteriors  in a direction more or less perpendicular to the $\sigma_8-\Omega_m$ degeneracy direction, suggesting that  the simultaneous  measurement of two- and three-point statistics can  be used to detect strong baryonic feedback.  The stronger the feedback, the  more gas is removed from the inner part of the haloes, leading to  larger differences between the posteriors for $\ave{M_{\rm ap}^2(\theta)}$  and  $\ave{M_{\rm ap}^3(\theta)}$. In particular,  one can see in the right panel of Figure \ref{fig:fisher} that strong feedback models, such as AGN, could already be tested, albeit at the $1\sigma$ level,  by measuring two- and three-point shear statistics  with the KiDS survey.
Note that the difference between this figure  and Figure 6 of Semboloni et al.\ (2011a) is  caused by the fact that we here use statistics and scales which are more sensitive to the baryonic feedback. In general dark matter only models are a bad description of baryonic  feedback but the amplitude of the bias in any parameter depends on the scales used to perform the cosmic shear analysis and on  the priors.  However, there is another factor which might cause small differences between  posteriors generated in this paper and Semboloni et al.\ (2011a): in that paper non-linear predictions were computed using the fitting formula of Smith et al. (2003) whereas here we used  the halo model. Both approaches suffer from limited accuracy and the cosmological parameters dependence is not the same. This issue has been  thoroughly discussed by Hearin et al. (2012) who also quantified the difference of the likelihood contours using different methods.

It is interesting to investigate whether the interpretation of  $\ave{M_{\rm ap}^3(\theta)}$  from the CFHTLS-Wide \cite{Hoekstraetal06,Fuetal08,Heymansetal12} 
dataset is already sensitive to  baryonic feedback. To mimic the CFHTLS-Wide dataset we assume the same source distribution  as {\it Euclid} ,  a survey  area  $A=170 ~{\rm deg^2}$, a galaxy density $n=15~{\rm gal/arcmin^2} $ and an intrinsic ellipticity dispersion $\sigma_e=0.3$ per component.  Under these assumptions, we find that  $\ave{M_{\rm ap}^2(\theta)}$  may be biased at the $1\,\sigma$ level  in the case of  strong feedback, such as the AGN scenario, and if all cosmological parameters but $\Omega_m$ and $\sigma_8$ are fixed. However, the statistical uncertainty for  $\ave{M_{\rm ap}^3(\theta)}$  is too large for the two likelihood posteriors to be in tension. 
Finally, we note  that  $\ave{M_{\rm ap}^3(\theta)}$ does not capture  all the information contained in the three-point shear statistic \cite{ScKiLo05}.  For example, by  measuring  $\ave{M_{\rm ap}(\theta_1) M_{\rm ap}(\theta_2) M_{\rm ap}(\theta_3)} $ one is able to capture better the statistical power of the three-point shear statistics.  We did not include such measurements here  as we have only a few lines-of-sight to estimate the covariance matrix. We expect that this  would increase the discrepancy between two- and three-point results by reducing the errors on the posteriors  of $\sigma_8$ and $\Omega_m$ from three-point shear statistics.   This might also be the case if one uses tomography.

\subsection{Calibration of baryonic feedback }\label{subsec:fit}

\begin{figure*}
\begin{tabular}{|@{}l@{}|@{}l@{}|}
\psfig{figure=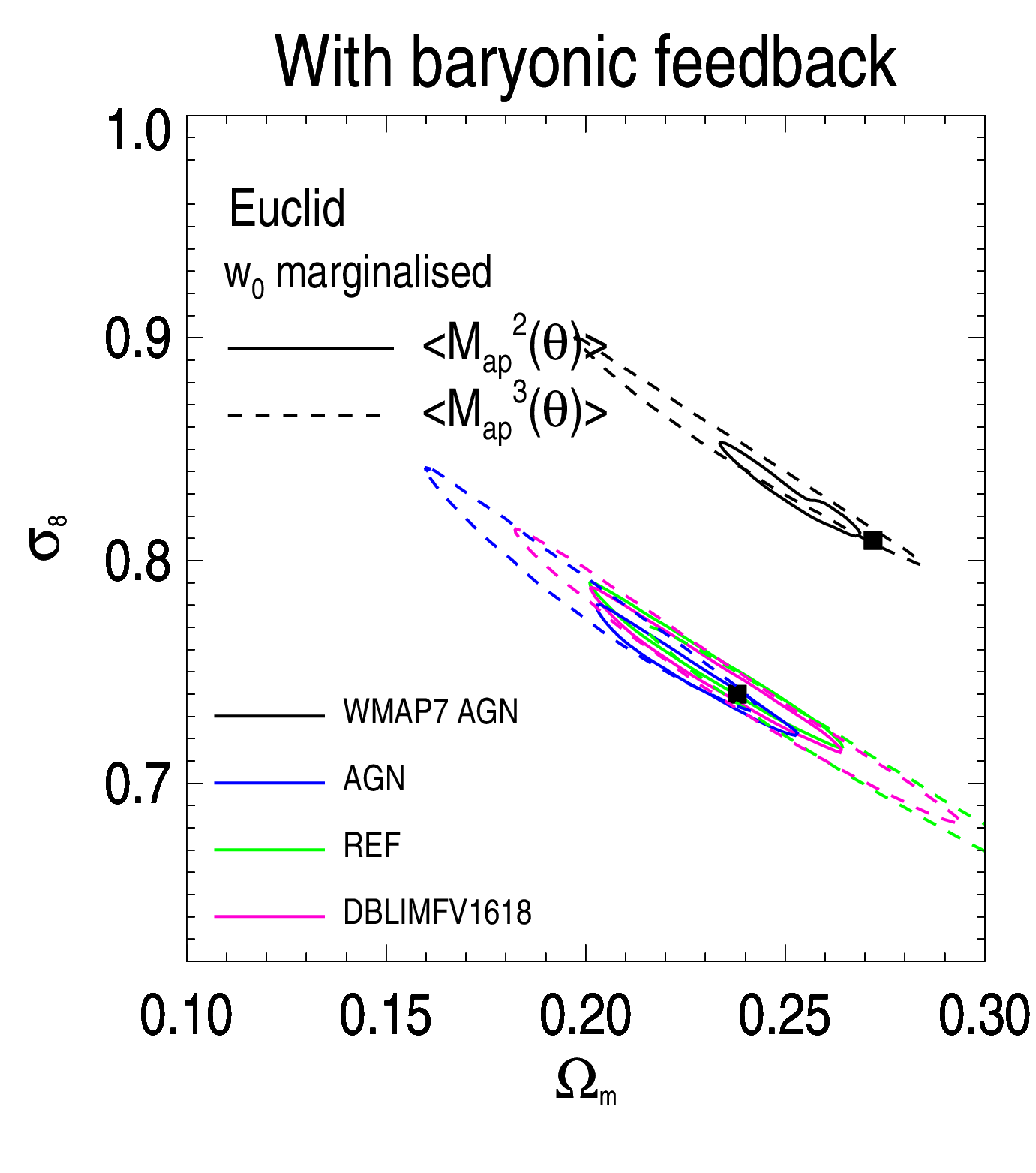,width=.45\textwidth}&
\psfig{figure=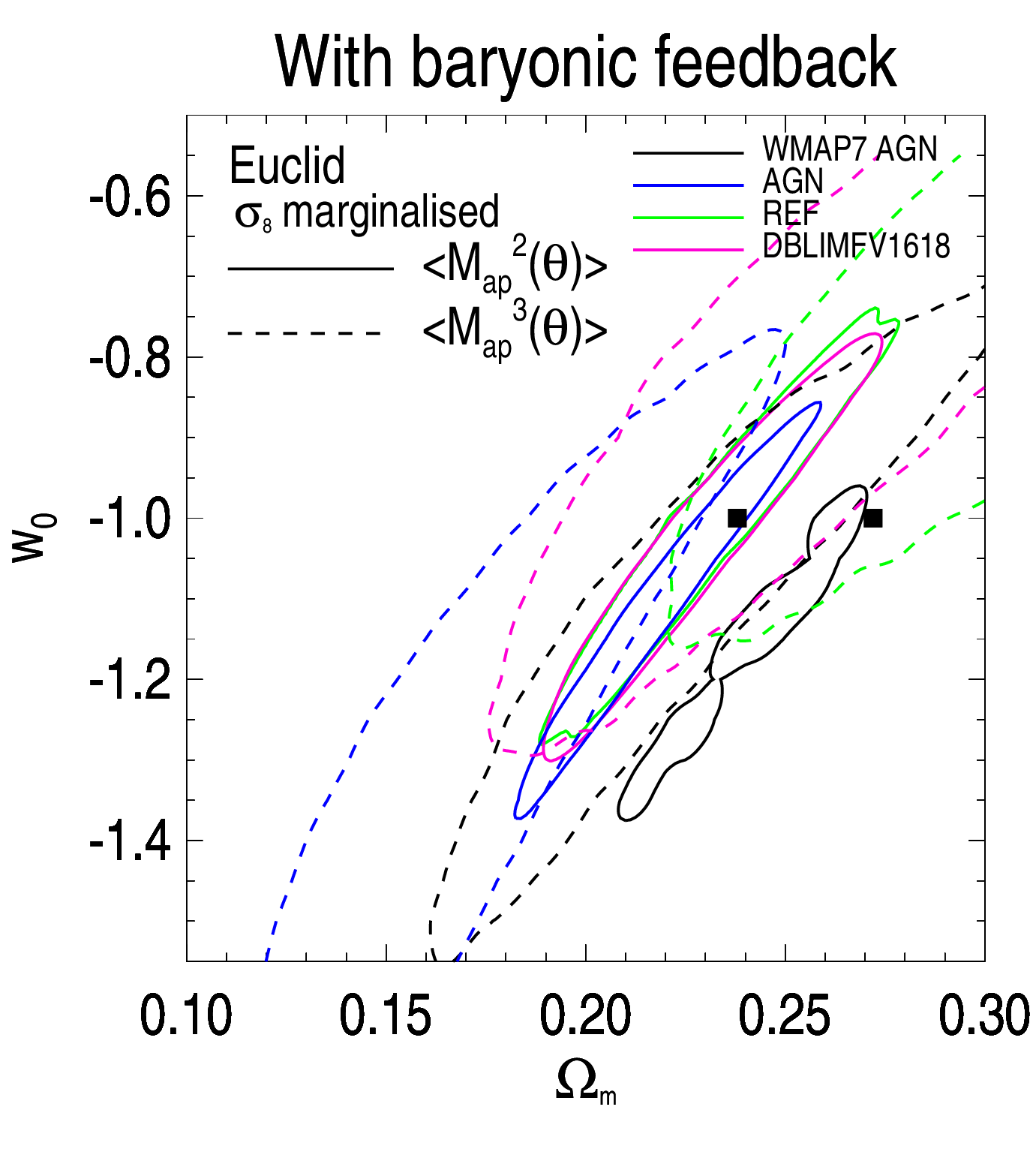,width=.45\textwidth}\\
\psfig{figure=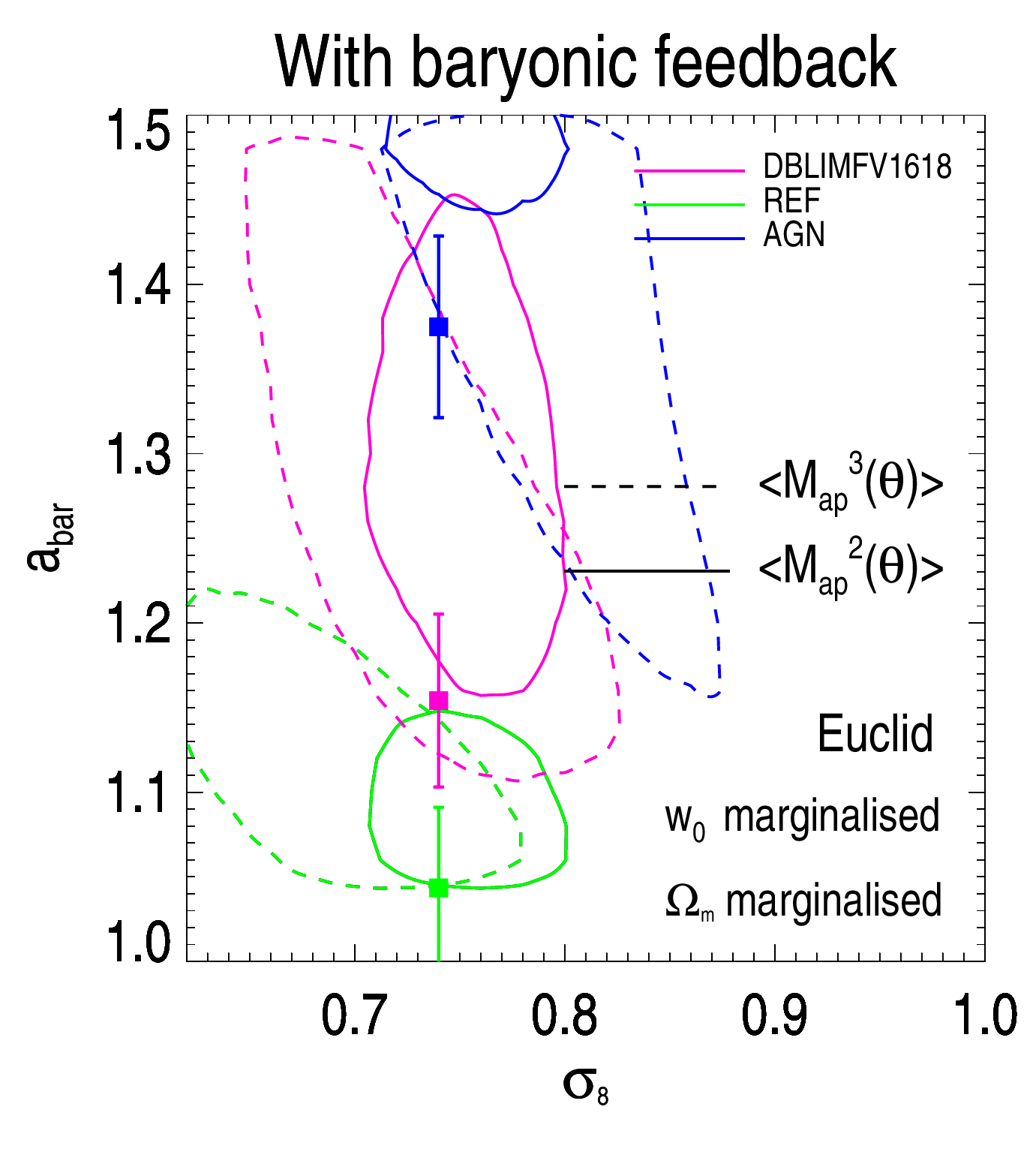,width=.45\textwidth}&
\psfig{figure=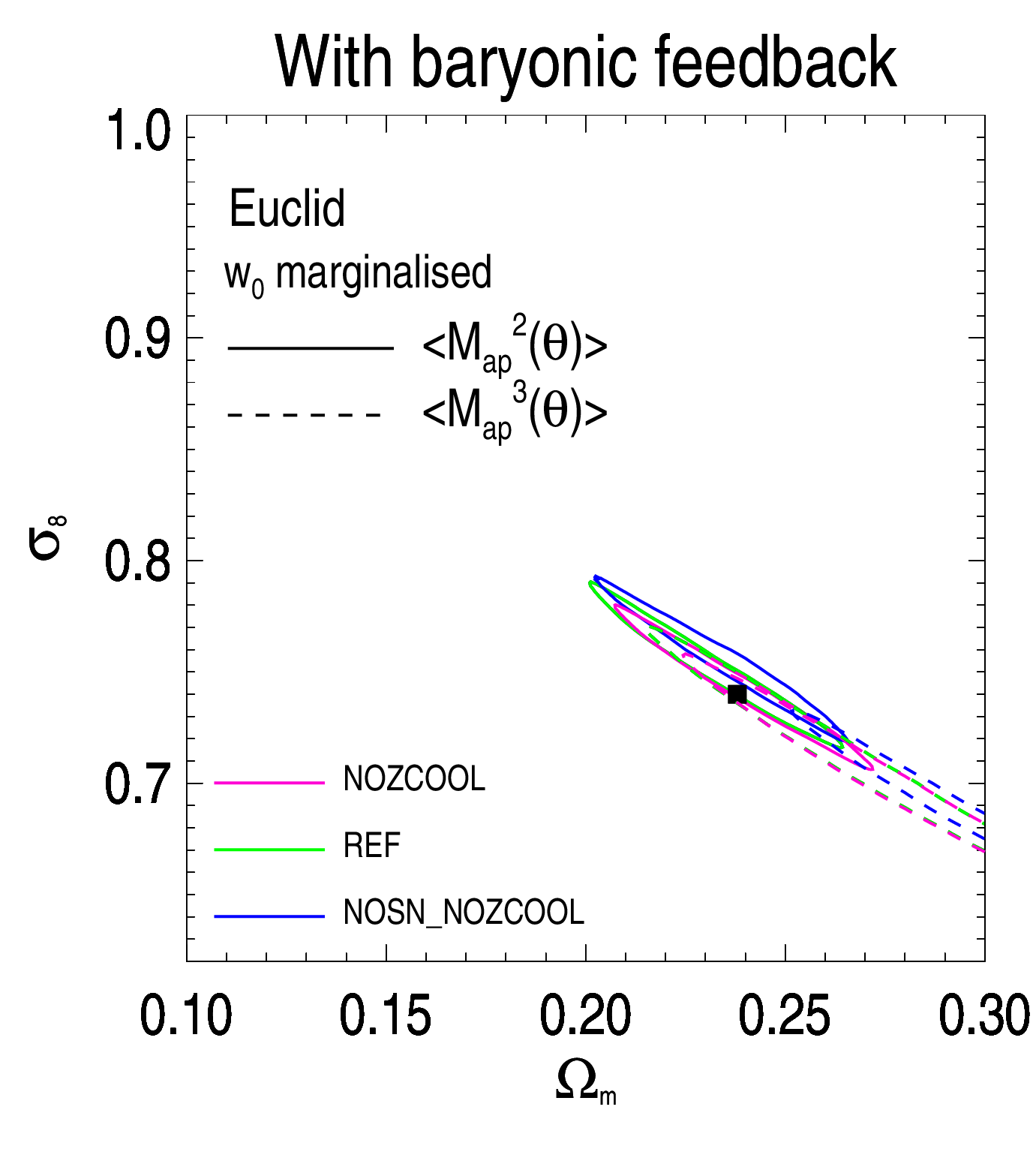,width=.45\textwidth}\\
\end{tabular}
\caption{\label{fig:likelihood_feedback}  The top-left (top-right) panel shows the probability distribution in the $\Omega_m$-$\sigma_8$ ($\Omega_m$-$w_0$) parameter space marginalised over $w_0$  ($\sigma_8$) and over the feedback parameters $\alpha$, $\beta$ and $a_{\rm bar}$. The results  are displayed for a survey similar to {\it Euclid}. Solid lines represent the  1 $\sigma$ contours inferred from $\ave{M_{\rm ap}^2(\theta)}$, while the dashed lines represent the same contours but inferred from $\ave{M_{\rm ap}^3(\theta)}$. In both panels, black squares indicate the value of the parameters of the input cosmology  which is the WMAP3 best-fit model for the AGN, REF and DBLIMFV1618 cases and the WMAP7 best-fit model for the WMAP7 AGN case.   
Bottom-left panel: posterior probability distribution for $a_{\rm bar}$ and $\sigma_8$  after marginalisation over $\alpha$, $\beta$, $\Omega_m$ and $w_0$. For each scenario, the average value of $a_{\rm bar}$ and its uncertainty derived in Section \ref{sec:newmodel} are also plotted. 
Bottom-right panel: same results as the top-left panel, but for the models REF, NOZCOOL and NOSN\_NOZCOOL, which were discussed in Section \ref{sec:newmodel}. }
\end{figure*} 
We have shown in the previous section that the occurrence of  baryonic feedback can be unveiled by a simple likelihood analysis of two- and three-point statistics measured on small scales where the effect is expected to be largest. In the case of strong feedback a dark matter only interpretation will lead to a discrepancy of the posterior distributions inferred from the two types of statistics. 

We have also shown  in Section \ref{sec:newmodel} that the effects of energetic feedback in each of the simulations can be described by a generic model that  allows us to parametrise the strength of the baryonic feedback irrespective of  the underlying physics that is responsible for removing the gas from the central parts of the haloes. Moreover, we have argued that there is a relation between the gas content and the stellar content in the simulations. This allowed us to parametrise the strength of the feedback by a single parameter $a_{\rm bar}$ which changes  from one scenario to another.

We will now use this parametrisation to study whether it is able to describe a general feedback model in such a way  that the final cosmological analysis for a generic unknown feedback scenario will be unbiased.  We vary the parameter $a_{\rm bar}$  in Equation (\ref{eq:fgas}) between the  extreme values measured in the OWLS simulations.
For a given $a_{\rm bar}$, we will assign the baryonic content and the stellar content to the haloes according to the best-fit relations shown in Figure \ref{fig:relations}.

As we already discussed, the mass-concentration  and the gas profile parameters  also depend on the feedback scenario (and on the cosmology). Since these parameters should be degenerate, we fix the mass-concentration relations to the one of the AGN simulation and keep $\alpha$ and $\beta$ as free parameters. In this way we have three nuisance parameter plus three cosmological parameters.

As we previously mentioned, it has it has been shown by van Daalen et al.\ (2011) that the change from a WMAP3 to a WMAP7 cosmology does not change the suppression of power caused by the baryonic feedback. We will assume here that this is also the case for other changes in the cosmological parameters, although we have shown that our halo model contradicts this. We have argued that this  can be explained by the fact that the dark matter profiles we use when we change cosmology are fixed to the ones obtained for the WMAP5 cosmology plus  feedback modifications derived within the WMAP3 cosmology. This is an important limitation of the current model but it could be overcome when more simulations are available.  To circumvent this problem we compute the modifications  for a WMAP3 cosmology  and we assume that these are the same for any other cosmology, by computing:
\be\label{eq:power_feedback}
P(k,z)^{\rm FB}=P(k,z)^{\rm DM}\times \frac{P(k,z)^{\rm FB}_{\rm WMAP3}}{P(k,z)^{\rm DM}_{\rm WMAP3}}
\ee
to obtain $P(k,z)^{\rm FB}$, the estimate of the power spectrum for a given cosmology and  feedback model.
Note that in this  way $\alpha$, $\beta$ and $a_{\rm bar}$ become nuisance parameters that span the  possible range of modifications of the power spectrum generated by baryonic feedback, but do not necessarily correspond to a physical quantity in a cosmology other than WMAP3.  Thus, the parameter space  now  consists of three cosmological parameters $\Omega_m,\sigma_8$, $w_0$, and  three nuisance parameters  $\alpha$, $\beta$, $a_{\rm bar}$.
We include among the possible scenarios the WMAP7 AGN for which we also derive the posterior probability.

Note that given the small number of simulations, we refrain  from combining  the likelihood analysis of two and tree-point  shear statistics. Instead we would like to show that our model is good enough to reduce the bias on the cosmological parameters for both   statistics. 
 In the top-left  panel of Figure \ref{fig:likelihood_feedback} we show constraints in the $\Omega_m$-$\sigma_8$  parameter space for the  REF, DBLIMFV1618, and AGN scenarios  and  also for the case of a WMAP7 cosmology with AGN feedback.  They are obtained after marginalising over the nuisance parameters   $\alpha$, $\beta$ and $a_{\rm bar}$ and over $w_0$.  The bias on the posterior is now comparable to the statistical error, also for the WMAP7 cosmology. We observe similar results for the $\Omega_m$-$w_0$  posterior probabilities,  displayed in the top-right panel of Figure \ref{fig:likelihood_feedback}. These results are obtained by marginalising  over the three nuisance parameters and over $\sigma_8$.

 In the bottom-left panel of Figure \ref{fig:likelihood_feedback} we show the posterior for the parameters $\sigma_8$-$a_{\rm bar}$.  These contours  have been obtained by marginalising over $\Omega_m$, $w_0$, $\alpha$ and $\beta$. They are compared with the $a_{\rm bar}$ values obtained from the fit performed in Section \ref{sec:newmodel}. One can see that there are differences between these values and the posterior probability  from the likelihood analysis. The differences increase with the strength of the feedback. This happens because, as  was already shown in   Figure \ref{fig:model_cosmo}, the modelling is not yet perfect. Thus, we expected to have some residual bias. We emphasise, however, that the marginalisation over  the parameters $\alpha$, $\beta$ and $a_{\rm bar}$  drastically  reduces the bias on the cosmological parameters, which are the ones we are interested in. 
Moreover,  we remind the reader that in this analysis  we used a maximum angular scale   $\theta=20\, {\rm arcmin}$ but that current surveys already allow measurements  on angular scales larger than a few degrees. Thus, the cosmological constraints are generally derived using measurements up to much larger scales than the ones we used here. Since at large scales the effect of the feedback is negligible, dark matter models perform well and the overall bias is reduced when these scales are included.  Note that, by construction, at large angular scales our feedback model reduces to the dark matter one. As one can see in Figure \ref{fig:map_hod}, this happens already for $\theta \lesssim 20~ {\rm arcmin}$. 

 In  the bottom-right panel of  Figure \ref{fig:likelihood_feedback}  we compare the posterior probability for the NOZCOOL and NOSN\_NOZCOOL models that do not follow the same empirical law we have applied to establish the fraction of gas and stars in our model (see Section \ref{sec:newmodel}).  As for the top-left panel, these results show constraints in the $\Omega_m$-$\sigma_8$  subspace after marginalising over the nuisance parameters   $\alpha$, $\beta$ and $a_{\rm bar}$ and over $w_0$.
The posteriors indicate only  a small bias,  comparable to the statistical errors,  even though our model is  wrong. The reason for this is that these models implement a weak feedback (i.e. the fraction of gas expelled outside the cores of the haloes is small)  which  changes the power spectrum mostly at small scales. Because of that, the uncorrected bias  is  small, similar to the one  obtained for the  REF scenario.  Marginalising over the feedback parameters, even using the wrong model, reduces this bias to within the $1\sigma$ level for NOZCOOL, whereas  the bias remains larger for the NOSN\_NOZCOOL scenario.                      

\section{Conclusions}\label{sec:conclusions}

Measurements of cosmic shear from upcoming weak lensing surveys have the potential to provide unique constraints on cosmological parameters and the theory of gravity. To interpret these measurements, models are required that can predict the signal with great accuracy. Van Daalen et al.\ (2011) have used simulations from the OWLS project (Schaye et al.\ 2010) to show that the baryonic physics of galaxy formation, in particular the strong feedback processes that are required to reproduce X-ray and optical observations of groups of galaxies, modifies the distribution of matter on such large scales that models will somehow have to account for these rather uncertain effects. Indeed, in Semboloni et al.\ (2011a) we demonstrated that cosmological  constraints based on two-point shear statistics will be  strongly biased if they are derived within a dark matter only framework.

Here we extended the analysis of Semboloni et al.\ (2011a) to three-point statistics. For a suite of OWLS models, we computed the three-point shear statistic $\ave{M_{\rm ap}^3(\theta)}$ using an approximate model based on perturbation theory (Scoccimarro \& Couchman 2001) that takes the non-linear power spectra tabulated by van Daalen et al.\ (2011) as input. 
We
 found that scenarios without strong feedback, which therefore suffer from the overcooling that has plagued most models in the literature, predict values of $\ave{M_{\rm ap}^3(\theta)}$ that agree with the dark matter only simulation to within a few per cent. However, simulations that include either strong stellar feedback (because of a top-heavy stellar initial mass function in starbursts) or AGN feedback, predict strong deviations from the dark matter only case on scales $\theta \la 10'$. In the case of AGN feedback, which is the only scenario that reproduces observations of groups of galaxies (McCarthy et al.\ 2010, 2011), the value of  $\ave{M_{\rm ap}^3(\theta)}$ is suppressed by up to $30-40\%$ on scales of a few arcmin. For comparison, the suppression measured for $\ave{M_{\rm ap}^2(\theta)}$ is about $20\%$ on the same angular scales. Hence, three-point statistics are even more affected by baryonic feedback than two-point statistics.

Because of the enhanced sensitivity to non-linear structure and the different cosmology dependence, ignoring baryonic processes in the interpretation of three-point shear statistics results in a bias that differs from the bias affecting two-point shear statistics. As a consequence, the posterior probability distributions for cosmological parameters are different for the two types of measurements.  Thus, the combination of two- and three-point shear statistics enables tests of galaxy formation models and allows one to check whether cosmological constraints inferred from one type of measurement may  be biased due to baryon physics. We find that this tension may already be detectable at the $\sim 1\sigma$ level using upcoming KiDS data and that it will be highly significant for the interpretation of {\it Euclid} data.  These cosmic shear measurements can be used along with more direct observations of the gas distribution in groups and clusters, including the outskirts, based on X-ray and Sunyaev-Zeldovich data, as well as galaxy-galaxy lensing observations. 
This is important, since our lack of knowledge of the physical mechanisms governing the overall distribution of baryons makes it impossible to construct simulations that can robustly predict the density field from first principles.

We presented a modified version of the halo model of Semboloni et al.\ (2011a) and used it to predict the non-linear matter power spectrum. The model captures the effects of baryonic feedback by introducing six parameters which characterise the dependence on halo mass of the hot gas and the stellar mass fractions as well as the shape of the gas profiles. All these parameters can in principle be measured from observations, but here we determined their values by fitting to the results of the simulations (Duffy et al.\ 2010; McCarthy et al.\ 2010, 2011; Semboloni et al.\ 2011a). For each scenario explored here, the model is able to reproduce the cosmic shear signal with an accuracy similar to that with which the standard halo model reproduces the dark matter only power spectra. 

As the accuracy with which the standard halo model can reproduce the dark
matter only predictions is insufficient for future surveys, and
because solving this issue is not the aim of our study, we corrected the
predictions of our model for the discrepancy between the halo model
and the OWLS dark matter only simulation. This correction also
mitigates any issues related to cosmic variance due to the finite size
of the simulation volume.

We found that although different scenarios predict different halo gas and stellar fractions, the predictions of all simulations that include both metal-line cooling and at least some energetic feedback follow the same relations between the amounts of hot gas and stars within a halo of a given mass and its baryon fraction. We used these relations, which are common to all physically plausible scenarios that we investigated, to reduce the number of parameters needed to describe the effects of feedback.  

We used this to  perform  likelihood analyses on simulated data, using a version of our halo model that contains three nuisance parameters to describe the effect of feedback on the matter distribution. Using this model, and marginalising over the three baryon-related parameters as well as over either the equation of state of the dark energy or $\sigma_8$, we found that the posterior probabilities for the cosmological parameters $\Omega_m$ and $\sigma_8$, or $w_0$ and $\Omega_m$,  based on measurements of $\ave{M_{\rm ap}^2(\theta)}$ and $\ave{M_{\rm ap}^3(\theta)}$ on scales between 0.5 and 20 arcmin are no longer discrepant, even for a survey as large as {\it Euclid}. With this approach the bias on the cosmological parameters became similar to, or smaller than the $1\sigma$
confidence regions while the overall statistical power was degraded only
mildly. Since cosmological constraints are generally obtained from data including larger scales, we expect the bias to be even
less significant, and the same should be true for the loss of information caused by
the presence of three extra nuisance parameters. 

One hypothesis underlying the use of our modified halo model, is that the feedback modifications are independent of the cosmology. This hypothesis is supported by the comparison between the WMAP3 and WMAP7 cosmologies reported by van Daalen et al.\ (2011), but it needs to be tested further using a wider range of cosmologies and ideally also using simulations that include modifications of gravity. The method also assumes that our model and the assumed relations between the baryonic mass fraction and the mass fraction of stars and hot gas are sufficiently flexible to describe the general modifications of the dark matter only predictions expected in realistic scenarios for baryonic feedback. This assumption also needs further testing using hydrodynamical simulations with a wider range of feedback implementations.

 Another limiting factor of this study  is the assumption that  the bispectrum can be written as a function of the power spectrum using a perturbative formula which has been derived for CDM  cosmologies. In the future, this approximate formula should be corrected  to account for any baryonic effects. 

Although the
performance of the model presented in this paper is still limited, and many of the underlying hypotheses remain to be tested,  it represents a
significant improvement over methods that rely solely on dark matter only predictions. In the future
the assumptions made in this paper will be investigated further and the model  will be better calibrated using simulations and available observations.

 %This  is also a main concern for the halo model itself which   does not reach  yet  the level of accuracy needed for the next generation of weak lensing measurements. Improving the performances of the halo model to reproduce the signal from dark-matter only simulations seems also to be a key ingredient.  In fact,  the halo model has the advantage compared to fitting formula to be constructed using  quantities which have a physical interpretation and because of that it offers an natural way  to correct a dark-matter only model for a feedback adding only  few other parameters which describe the matter distribution inside halos independently of the physics behind the feedback model. For this reason hybrid approaches which try to conciliate perturbation theory and halo model are very promising \cite{VaNi11}.

\section*{Acknowledgments}
We are grateful to Marcel van Daalen, Marco Velliscig, Ian McCarthy, Alan Duffy and all other members of the OWLS collaboration for discussions, for making the OWLS project possible and for their efforts in making the simulation measurements that were used here.  We thank Cosimo Fedeli, Andrew Zentner, Scott Dodelson, Elisabeth Krause  and Tim Eifler  for discussions that stimulated the progress of this project. We are also grateful to Catherine Heymans, Ludo van Waerbeke and Sanaz Vafaei for providing us with the simulations used to compute the covariance matrices needed in this paper. 
ES, HH and JS acknowledge support from the Netherlands Organization for Scientific Research (NWO) through VIDI grants. ES and HH also  acknowledge the support from the ERC under FP7 grant number 279396. The research leading to these results has received funding from the European Research
Council under the European Union's Seventh Framework Programme (FP7/2007-2013) / ERC Grant agreements 278594-GasAroundGalaxies and from the Marie Curie Training Network CosmoComp (PITN-GA-2009-238356).
This material is based in part upon work supported in part by the National Science Foundation Grant No.\ 1066293 and the hospitality of the Aspen Center for Physics.

\end{document}